\documentclass[aps,prl,twocolumn,superscriptaddress,preprintnumbers,nofootinbib]{revtex4-1}
\usepackage{amsmath,amssymb,graphicx,natbib,aas_macros,bm}
\usepackage{enumerate}
\usepackage{hyperref}

\usepackage{color}

\newcommand{\be}{\begin{equation}}
\newcommand{\ee}{\end{equation}}
\newcommand{\bea}{\begin{eqnarray}}
\newcommand{\eea}{\end{eqnarray}}

\begin{document} 


\title{Singlet Assisted Vacuum Stability and the Higgs to Diphoton Rate}

\author{Brian Batell}
\affiliation{Enrico Fermi Institute and Department of Physics, University of Chicago, Chicago, IL,
60637}

\author{Sunghoon Jung}
\affiliation{School of Physics, KIAS, Seoul 130-722, Korea}

\author{Hyun Min Lee}
\affiliation{School of Physics, KIAS, Seoul 130-722, Korea}

\begin{abstract}
Light, electrically charged vector-like `leptons' with ${\cal O}(1)$ Yukawa couplings can enhance the $h\rightarrow \gamma\gamma$ rate, as is suggested by measurements of the signal strength $\mu_{\gamma\gamma}$  by ATLAS and CMS. 
However, the large Yukawa interactions tend to drive the Higgs quartic coupling negative at a low scale $\Lambda \lesssim 10$ TeV, 
and as such new physics is required to stabilize the electroweak vacuum. A plausible option, which does not rely on supersymmetry, 
is that the Higgs in fact has a much larger tree level quartic coupling than in the Standard Model, which is possible with an extended scalar sector. 
We investigate in detail the minimal model with a new real singlet scalar which condenses and mixes with the Higgs.
The mechanism is very efficient when the singlet is heavier than the weak scale $m_s \gtrsim$ TeV, in which case  the vacuum can be stable and the couplings can be perturbative up to scales $\Lambda \sim 10^9$ GeV while $\mu_{\gamma\gamma} \sim  1.7$. On the other hand, for a singlet at the weak scale, $m_s \sim v$, a substantial shift in the Higgs quartic coupling requires sizable Higgs-singlet mixing. Such mixing can be utilized to further enhance $\mu_{\gamma\gamma}$ through Yukawa couplings of the singlet to the vector-like leptons. In this case, for an enhancement  $\mu_{\gamma\gamma} \sim  1.7$ it is possible to obtain a stable vacuum up to scales $\Lambda \sim 1000$ TeV.  
The singlet can have significant mixing with the Higgs particle, allowing for its resonant production via gluon fusion at the LHC, and can be searched for through Higgs like decays and decays to vector-like leptons, which can lead to multi-lepton final states.  We also 
comment on UV extensions of the minimal model. 
\end{abstract}

\maketitle

A new Higgs-like particle has been discovered at the LHC~\cite{ATLAS-Higgs, CMS-Higgs}, 
potentially putting into place the final piece of the Standard Model (SM). The determination of the properties of this particle is absolutely paramount, as any deviations from those of the SM Higgs could signal the presence of new physics (NP). In this regard, it is of interest to examine the 
measurements of the Higgs  signal strength modifier for the channel $i$, defined as 
\begin{equation}
\mu_i = \frac{ \sigma(pp\rightarrow h\rightarrow i)  }{ \sigma(pp\rightarrow h\rightarrow i)_{\rm SM}    }.
\label{signal}
\end{equation}
At this early stage of data taking  the $\gamma\gamma$, $ZZ$, and $WW$ channels have the best statistical sensitivity. 
In particular, naively combining the results of the ATLAS, CMS, and Tevatron collaborations~\cite{ATLAS-Higgs, CMS-Higgs,:2012zzl}, we obtain the values
\begin{equation}
\mu_{\gamma\gamma} = 1.7\pm0.4,~~~~\mu_{VV} = 0.8\pm0.3,~~~~
\label{pattern}
\end{equation}
where $VV = WW,ZZ$.  One observes that the $ZZ$ and $WW$ channels are in agreement with the SM expectations while the $\gamma\gamma$ channel displays a slight enhancement.

Of course, the deviation in the $\gamma\gamma$ channel is not statistically significant at this time and may very well diminish as more data are analyzed. However, we have entered the era of precision Higgs physics, and in this light it is certainly worthwhile to consider the implications of modified Higgs couplings for physics beyond the SM. Taking the pattern (\ref{pattern}) at face value, the simplest and most plausible NP explanation is that the enhancement in $\mu_{\gamma\gamma}$ is due to new color-singlet, electrically charged particles that couple to the Higgs boson. 
Such states provide an additional one loop contribution to the decay 
 $h\rightarrow \gamma\gamma$ while not significantly altering the couplings of the Higgs to other SM particles. This has been pointed out after the December 2011 hint of a 125 GeV excess in Refs.~\cite{Carena:2011aa,Batell:2011pz}, and has also been explored many times since.
 
In this paper we will consider the possibility that the new particles are color-singlet, electrically charged vector-like `leptons'. 
Such fermions require ${\cal O}(1)$ Yukawa couplings in order to bring about a large enhancement in $\mu_{\gamma\gamma}$~\cite{VLfermion,Joglekar:2012vc,ArkaniHamed:2012kq}. 
However, as emphasized recently by Refs.~ \cite{Joglekar:2012vc,ArkaniHamed:2012kq,Reece:2012gi,lpp2}, these fermions yield a large negative contribution to the $\beta$ function of the Higgs quartic coupling. The Higgs quartic coupling is thus driven to negative values at a very low scale $\Lambda \lesssim 10$ TeV, threatening the stability of the electroweak vacuum. Thus, the simplest models designed to enhance the $h\rightarrow \gamma\gamma$ rate with new charged fermions require a UV completion at a very low scale.  

One obvious solution is to supersymmetrize these theories, in which case the scalar superpartner of the new fermion will provide a compensating contribution to the Higgs quartic $\beta$ function. Another possibility, which is the focus of this paper, is that the Higgs quartic coupling $\lambda_H$ could in fact be much larger than it is in the SM. This is possible if nature has chosen a more complicated scalar sector than the minimal SM. Indeed, if the Higgs mass mixes with another scalar field that obtains a vacuum expectation value (VEV), then the quartic coupling of the Higgs field can be significantly larger than its SM value. If so, the scale at which NP is required to stabilize the vacuum can be pushed to much higher scales, as we will demonstrate below. In this paper we will investigate the minimal model in which this mechanism can be applied, which contains a new real scalar singlet field. 

The basic mechanism to stabilize the vacuum investigated here has already been presented in Ref.~\cite{EliasMiro:2012ay,lebedev} and was applied to the vacuum stability of the SM. In that case, since the Higgs quartic coupling runs very slowly in the SM, the singlet scalar could be much heavier than the weak scale and still stabilize the vacuum. However in our case, since the new vector-like leptons cause the Higgs quartic coupling to run negative very quickly, the singlet scalar must be comparatively light to the case of Ref.~\cite{EliasMiro:2012ay}, and may even reside at the weak scale, in which case it can be searched for at the LHC.

Indeed, if the singlet mass is around the weak scale, a sizable mixing with the SM Higgs is required in order to cause a large positive shift in the Higgs quartic coupling. Because the singlet scalar mixes with the Higgs, it picks up a coupling to other SM particles. This allows it to be produced resonantly at the LHC via gluon fusion. The singlet resonance can be detected by searching for Higgs-like decay modes or decays to pairs of vector-like leptons.

\section{ Enhanced $h\rightarrow \gamma\gamma$ and Vacuum Stability}

We begin by briefly reviewing the vacuum stability issue for theories in which new electrically charged fermions enhance the $h\rightarrow \gamma\gamma$ rate\footnote{See Refs.~\cite{Dobrescu:2000jt,Draper:2012xt} for a novel mechanism to enhance the apparent diphoton signal utilizing Higgs decays to boosted multi-photon final states, and further related studies in Refs.~\cite{Toro:2012sv,Ellis:2012sd}.   }. For concreteness, consider two charged fermions with a mass matrix 
\begin{equation}
{\cal M} = \left(  
\begin{array}{cc}
M_1  & y_1 v/\sqrt{2} \\
y_2 v/\sqrt{2}  & M_2
\end{array}  \right),
\label{eq:basicmass}
\end{equation}
where $M_{1,2}$ are vector-like mass terms, $y_{1,2}$ are Yukawa couplings of the fermions to the Higgs field, and $v = 246$ GeV is the Higgs VEV.
The modifications to the Higgs-photon-photon coupling are well described via the low energy Higgs theorems~\cite{Ellis:1975ap},~\cite{Shifman:1979eb}:
\begin{eqnarray}
{\cal L} & \supset & \frac{\alpha}{16 \pi v}    b_{EM} \left(  \frac{\partial}{\partial \log v} \log \det {\cal M} {\cal M}^\dag \right)    h F_{\mu\nu} F^{\mu\nu} ,
\end{eqnarray}
 where the QED $\beta$ function coefficients are  $b_{EM} = (4/3) Q_f^2$. For the mass matrix of equation
(\ref{eq:basicmass}) we obtain
\begin{equation}\label{eq:zf}
  \frac{\partial}{\partial \log v} \log \det {\cal M} {\cal M}^\dag = -\frac{2 y_1 y_2 v^2}{M_1 M_2-y_1 y_2 v^2 /2},
\end{equation}
and the QED $\beta$ function coefficient is  $b_{EM} = (4/3) Q_f^2$.

The modification to the Higgs signal strength in the diphoton channel $\mu_{\gamma\gamma}$ (defined in Eq.~(\ref{signal})) in this setup arises solely through the new contribution to the $h\rightarrow \gamma\gamma$ partial width. Using the effective Lagrangian above, we obtain
\begin{eqnarray}
\label{rgam}
\mu_{\gamma \gamma}
& = & 
\frac{\Gamma(h\rightarrow \gamma \gamma)}{\Gamma(h\rightarrow \gamma \gamma)_{\rm SM}}  
 =  \bigg\vert 1+\frac{ b_{EM} }{A_\gamma^{\rm SM}} \frac{y_1 y_2 v^2}{M_1 M_2-y_1  y_2 v^2 /2}  \bigg\vert^2, \nonumber \\&&
 \label{ssLET}
\end{eqnarray}
with $A_\gamma^{\rm SM} \simeq 6.5$ is due to the competing $W$ boson and top quark contributions. We see that large ${\cal O}(1)$ Yukawa couplings are required to obtain an enhancement of the size indicated by the data (\ref{pattern}). As an illustration, assuming a charge $Q=1$, a common vector-like mass of $M_1 = M_2 = 300$ GeV, then Yukawa couplings of $y_1 =  y_2 = 1$ are required to obtain an enhancement of $\mu_{\gamma\gamma} \simeq 1.5$. 

\begin{figure}[t]
\centerline{
\includegraphics[width=0.45\textwidth]{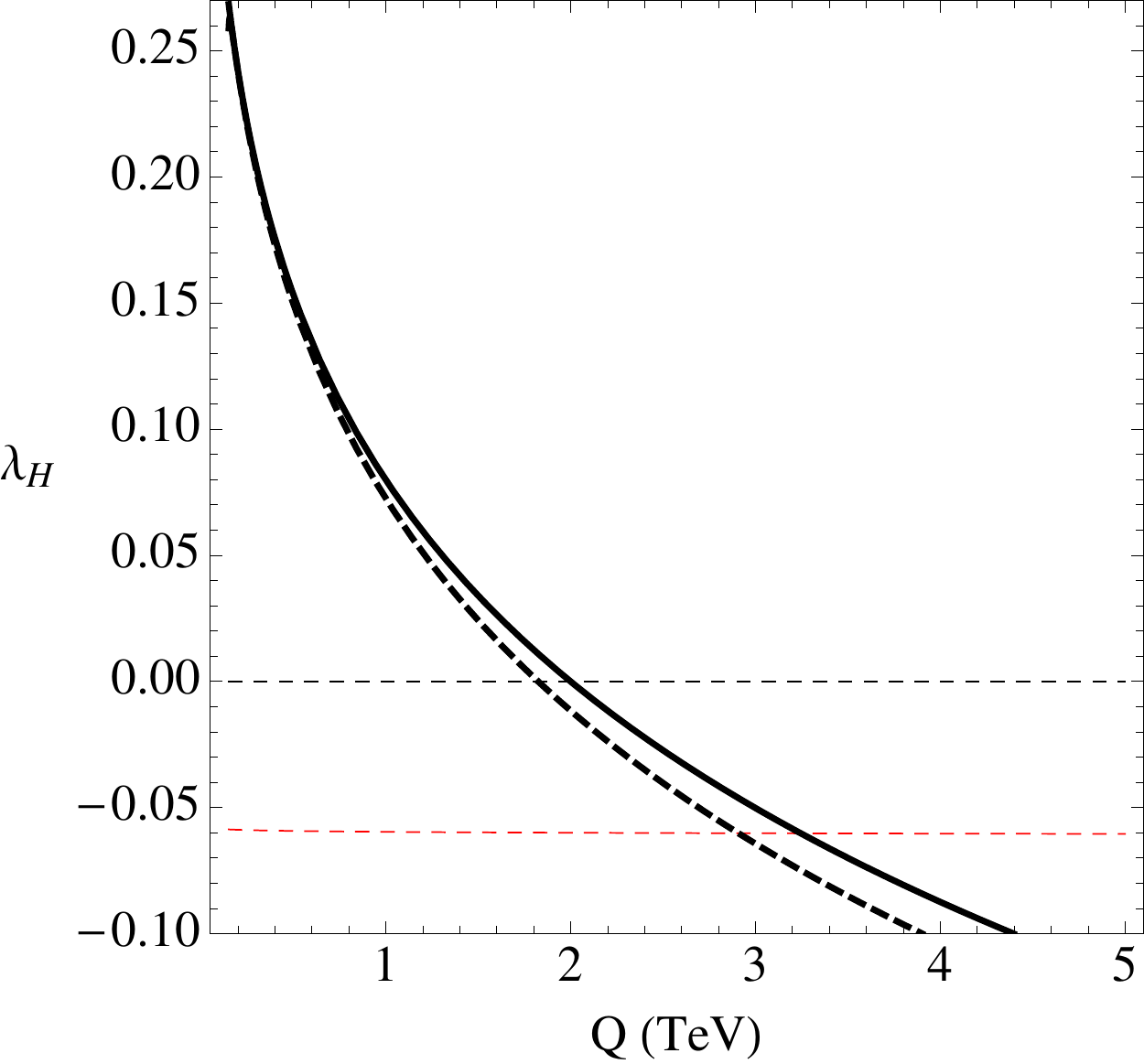}~~
}
\caption{\small Running of the Higgs quartic coupling $\lambda_H$ for $y_1 = y_2 = 1$. The solid line shows the exact numerical solution to the RG equation, while the dashed line shows the approximate analytic solution.  The coupling $\lambda_H$ runs negative at a low scale. }
\label{fig:quartic}
\end{figure}
However, the large Yukawa couplings give negative contributions to the $\beta$ function of the Higgs quartic coupling $\lambda_H$ proportional to $- y_{i}^4$,  causing $\lambda_H$ to run negative at a low scale and thus destabilizing the vacuum. 
To see this clearly, let us write the renormalization group (RG) equation for $\lambda_H$:
\begin{eqnarray}
16\pi^2 \frac{d\lambda_H}{dt} & = &  12 \lambda_H^2 + a  \lambda_H + b, \label{RGapprox}\\
  a & =  &  12 y_t^2+ 4(y_1^2+ y_2^2 ) +\dots ,    \\
  b &  = &   -12 y_t^4- 4(y_1^4+ y_2^4) 
  +\dots   , 
\end{eqnarray}
where $t = \log Q /Q_0$, with $Q$ the RG scale, and the ellipses denote sub-leading contributions from the gauge couplings.
The Yukawa couplings run very slowly compared to the Higgs quartic coupling, and thus $a$ and $b$ are approximately constant over the scales $Q$ of interest. 
Furthermore, for a 125 GeV Higgs boson $\lambda_H = m_h^2/v^2 \simeq 0.26$, so that we can neglect the small $\lambda_{H}^2$ term in the RG equation (\ref{RGapprox}). 
With these approximations, the solution to the RG equation can be obtained by integration and is given by
\begin{equation}
\lambda_H(Q)  = \left(\lambda_H(Q_0) + \frac{b}{a} \right) \left( \frac{Q}{Q_0}\right)^{a/16\pi^2} -  \frac{b}{a}.
\label{lamapprox}
\end{equation}
Compared to the SM, the additional fermions with Yukawa couplings $y_i$ in Eq.~(\ref{RGapprox}) lead to a larger exponent $a/16\pi^2$ causing more rapid variation with scale $Q$. Furthermore, note that for $y_i \gtrsim 1$, the competing negative term in (\ref{lamapprox}) proportional to $b/a$ becomes stronger, since $|b/a|$ is larger than in the SM, thus leading to negative values of $\lambda_H$ at lower scales $Q$. 
In Fig.~\ref{fig:quartic} we display the evolution of the quartic coupling $\lambda_H$ for Yukawa couplings $y_1 = y_2 = 1$ and $Q_0 = m_t$. The red dashed line indicates the metastability bound from Ref.~\cite{Isidori:2001bm}. The figure shows that the approximate solution agrees well with the exact one which we have obtained by numerically integrating the full RG equations\footnote{The full set of RG equations can be obtained from those presented in the Appendix by setting the additional couplings associated with the extended singlet sector to zero.}.

Inspection of the approximate solution in Eq.~(\ref{lamapprox}) also reveals two possible mechanisms to stabilize the vacuum to higher scales: 1) make $a, |b|$ smaller by adding new particles that couple to the Higgs field, as in {\it {\it e.g.}} a supersymmetric version of the model, and 2) make the Higgs quartic coupling $\lambda_H(Q_0)$ larger. We will now focus on the second possibility, which is possible if there is an extended scalar sector.

\section{Large Higgs Quartic from  an extended scalar sector}

We now consider the simplest model of an extended scalar sector that allows for a large Higgs quartic coupling. The model contains the Higgs field $H\sim (1,2,-1/2)$ along with a real singlet scalar field $S\sim (1,1,0)$.
The singlet scalar mixes with the Higgs particle, allowing for the Higgs quartic coupling $\lambda_H$ to be larger than its value in the SM, 
$\lambda_H^{\rm SM} = m_h^2/v^2 \simeq 0.26$, thus allowing the vacuum to be stable to much higher scales. Much of the discussion here follows that of Ref.~\cite{EliasMiro:2012ay}, to which we refer for more details.

We consider the following scalar potential: 
\begin{eqnarray}
V(H,S)&=&\frac{1}{2}\lambda_H (|H|^2-v^2/2)^2
 +\frac{1}{8} \lambda_S (S^2-w^2)^2 \nonumber \\
& + &  \frac{1}{2}\lambda_{HS} (|H|^2-v^2/2)(S^2-w^2).
\label{potential}
\end{eqnarray}
The scalar potential written above is invariant under a discrete $Z_2$ symmetry $S\rightarrow -S$. Below we will couple the singlet to the vector-like leptons responsible for enhancing the diphoton rate in the presence of the mixing with the Higgs, and this coupling will explicitly break the $Z_2$ symmetry. Radiative corrections will induce tadpole and cubic terms to the potential, but parametrically suppressed compared to the large tree level terms considered here. We will comment further on this below, but for now we examine the potential in Eq.~(\ref{potential}).

The scalar potential must be bounded from below, which imposes the restrictions
$\lambda_H >0$, $\lambda_S >0$, and, if $\lambda_{HS}$ is negative,  $\lambda_H \lambda_S > \lambda_{HS}^2$. 
If these conditions are satisfied, and $v^2>0$ and $w^2>0$, then 
both the Higgs and the singlet condense, $\langle H \rangle^T = (v/\sqrt{2}, 0)$,  
$\langle S \rangle = w$. Expanding about the vacuum in the fluctuations $\hat h, \hat s$, we obtain the mass matrix in the scalar sector $(\hat h, \hat s)$:
\begin{equation}
{\cal M}^2=\left(\begin{array}{ll} \lambda_H v^2 & \lambda_{HS} vw \\
\lambda_{HS} vw & \lambda_S w^2 \end{array}\right).
\label{scalars}
\end{equation}
There are two limits of the model to consider: 1)  the singlet is much heavier than the weak scale, $w \gg v$, and 2) the singlet  is close to the weak scale, $w\sim v$. 

\begin{figure}[t]
\centerline{
\includegraphics[width=0.45\textwidth]{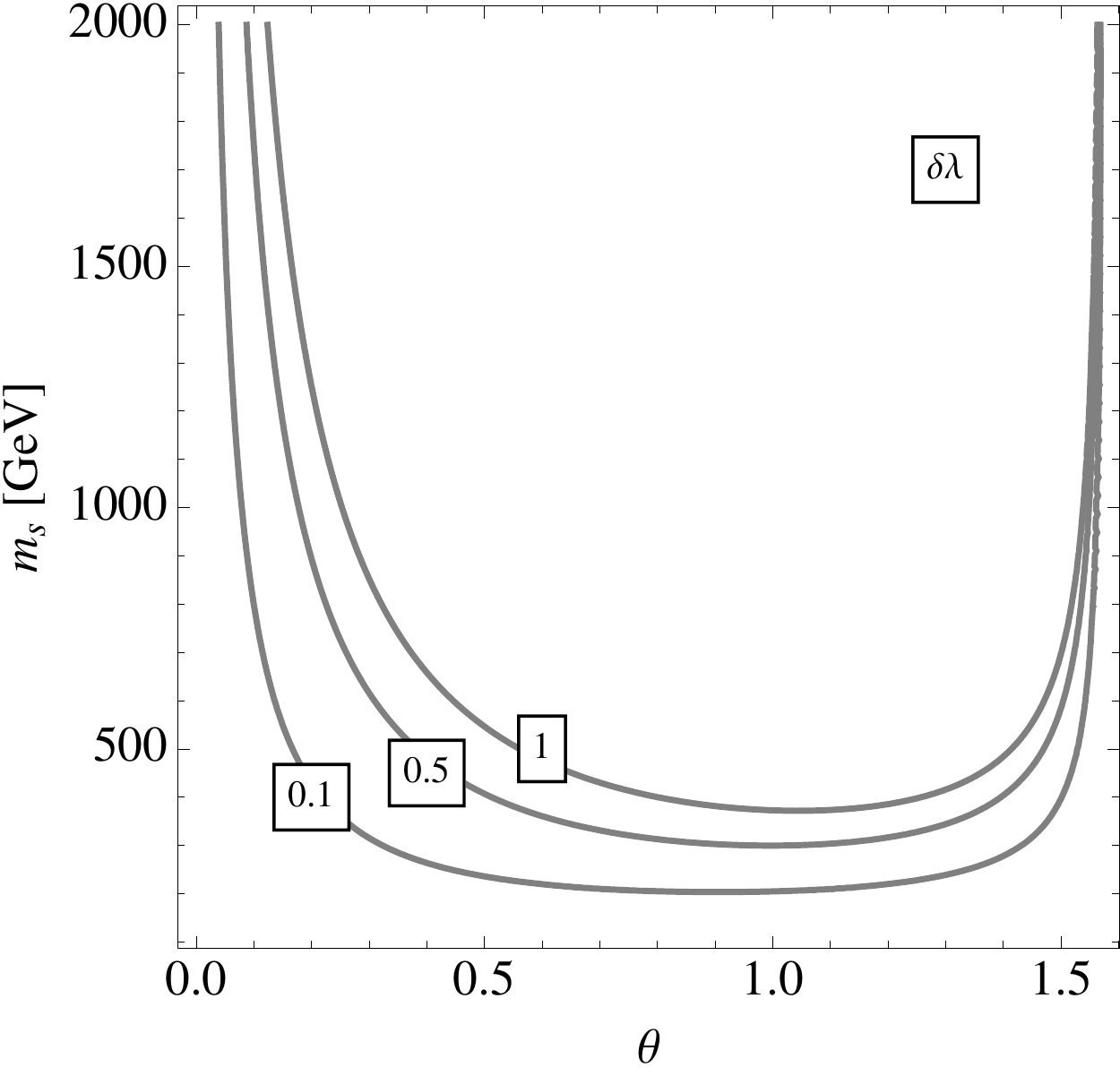}~~
}
\caption{\small Isocontours of the quartic coupling shift $\delta \lambda \equiv \lambda^2_{HS}/\lambda_S$ in the $m_s - \theta$ plane.}
\label{fig:delThM}
\end{figure}
Let us first consider the case of the heavy singlet. In the regime $Q < m_s \simeq \sqrt{ \lambda_S} w$ we may integrate out the singlet using the equations of motion for $S$, leading to the following effective potential for the Higgs field $H$:
\begin{equation}
V_{\rm eff}  =\frac{1}{2}\lambda_{\rm eff} (|H|^2-v^2/2)^2,
\label{effpot}
\end{equation}
where the effective Higgs quartic coupling is 
\begin{equation}
\lambda_{\rm eff} = \lambda_H -\frac{\lambda_{HS}^2}{\lambda_S}.
\label{effQ}
\end{equation}
It will be convenient to define
\begin{equation}
\delta \lambda \equiv \frac{\lambda_{HS}^2}{\lambda_S},
\label{shift}
\end{equation}
which is simply the tree level shift in the Higgs quartic coupling in Eq.~(\ref{effQ}) from the matching of the 
full theory to the effective theory at the scale $Q = m_s$.
For very reasonable choices of the quartic couplings $\lambda_S$, $\lambda_{HS}$ it is possible to obtain a large shift $\delta \lambda$ and thus a large Higgs quartic coupling
$\lambda_H$ in the full theory.

When the singlet is light, $m_s \sim v$, we should instead diagonalize the system (\ref{scalars}) via the orthogonal transformation
\begin{equation}
\left( 
\begin{array}{c}
\hat h \\
\hat s
\end{array}
\right) = 
\left( 
\begin{array}{cc}
c_\theta & s_\theta  \\
-s_\theta & c_\theta
\end{array}
\right)
\left( 
\begin{array}{c}
h \\
s
\end{array}
\right),
\end{equation}
where the mixing angle is defined via
\begin{equation}
\tan\,2\theta= \frac{2\lambda_{HS}vw}{\lambda_H v^2-\lambda_S w^2}.
\end{equation}
The masses of the physical scalars $h$ and $s$ are 
\begin{equation}
m^2_{h,s}\!=\!\frac{1}{2}\Big(\lambda_H v^2+\lambda_S w^2 \mp \sqrt{(\lambda_S w^2\!-\!\lambda_H v^2)^2\!+\!4\lambda^2_{HS} v^2 w^2}
\Big),
\end{equation}
The tree-level shift in the coupling $\delta \lambda$ in Eq.~(\ref{shift}) can be written explicitly in terms of the physical parameters as
\begin{equation}
\delta \lambda = \frac{(m_s^2- m_h^2) s^2_\theta c^2_\theta }{v^2 (m_s^2 c^2_\theta + m_h^2 s^2_\theta)}.
\end{equation}
We show in Fig.~\ref{fig:delThM} values of $\delta\lambda$ in the $\theta - m_s$ plane. One observes that for light singlets $m_s \sim v$, a large mixing between the Higgs and the singlet is required to obtain a large tree-level shift $\delta \lambda$. On the other hand, as $m_s$ becomes much heavier than the weak scale, large shifts can be obtained even for small mixing angles. Indeed, in the limit $m_s \gg m_h$, the mixing angle becomes $\theta \simeq \delta\lambda\, v / m_s$ showing that even for ${\cal O}(1)$  values of $\delta\lambda$ the mixing angle will be small.

Note that this mechanism relies on a cancellation between two ${\cal O}(1)$ quantities, and thus involves some degree of tuning. Clearly the tuning does not have to be severe to obtain the desired effect. For example, a quartic coupling 
$\lambda_H \sim {\cal O}(1) $, implying a cancellation at the  $\sim 10-50\%$  level, will allow for a stable vacuum to much higher scales, as we will demonstrate below.

\subsection{Incorporating vector-like leptons}

We now describe explicitly the vector-like lepton sector to be examined. 
The model contains the following Weyl fermion fields:
\begin{eqnarray}
l_4 = \left( 
\begin{array}{c}
\nu_4 \\
e_4
\end{array}
\right) \sim (1,2,-1/2), 
& e_4^c\sim (1,1,1), 
& \nu_4^c\sim (1,1,0),\nonumber \\
\tilde \l_4 = \left( 
\begin{array}{c}
\tilde e_4 \\
\tilde \nu_4
\end{array}
\right) \sim (1,\bar 2,1/2),~~
& \tilde e_4^c\sim (1,1,-1), 
&\,  \tilde \nu_4^c\sim (1,1,0). \nonumber \\ &&
\end{eqnarray}
The Yukawa couplings and vector-like mass terms are 
\begin{eqnarray}
-{\cal L} & = &  M_l l_4 \tilde l _4 + M_e {e_4^{c}} \tilde {e_4^c} + M_\nu {\nu_4^c} \tilde {\nu_4^c}   \nonumber \\
&+&  y_e H l_4{ e_4^c} + \tilde y_e \tilde l_4 H^\dag  \tilde {e_4^c} + 
  y_\nu H^\dag l_4 {\nu_4^c} + \tilde y_\nu  \tilde l_4 H \tilde {\nu_4^c} \nonumber \\
  & + & \frac{x_l}{\sqrt{2} } S l_4 \tilde l_4 + \frac{x_e}{\sqrt{2} }S {e_4^c} \tilde {e_4^c} + \frac{x_\nu}{\sqrt{2}} S {\nu_4^c} \tilde {\nu_4^c} +{\rm h.c.}\,.
  \label{eq:Lferm} 
 \end{eqnarray}
For simplicity we neglect possible Majorana mass terms for the neutral fermions and assume all parameters in Eq.~(\ref{eq:Lferm}) are real. However, see Refs.~\cite{Voloshin:2012tv,McKeen:2012av}  for recent studies of the effects of new CP-violating phases from vector-like fermions responsible for the diphoton enhancement. 

As alluded to above, the $Z_2$ symmetry of the classical scalar potential in Eq.~(\ref{potential}) is explicitly broken by the Yukawa couplings of the singlet to the vector-like leptons in Eq.~(\ref{eq:Lferm}), and as such, $Z_2$ breaking terms in the potential will be generated radiatively. The relevant terms are
\begin{equation}
\Delta V = \mu_1^3 S + \mu_3 S^3 + \mu_{HS} S |H|^2.
\end{equation}
Assuming the tree level couplings of these terms are small, the parametric size of these couplings coming from one-loop exchange of vector-like leptons is
\begin{eqnarray}
\mu_1^3 \sim \frac{x m^3}{16 \pi^2}, ~~~~  \mu_3 \sim   \frac{x^3 m}{16 \pi^2} ,~~~~
 \mu_{HS} \sim \frac{x y^2 m}{16 \pi^2}.
\end{eqnarray}
where we have used $x$, $y$, and $m$ as a generic label for the Yukawa couplings and fermion masses. 
For ${\cal O}(1)$ Yukawa couplings, and fermion masses near the weak scale, the size of these couplings is  small compared to the effects that will be induced by the tree level $Z_2$ symmetric potential (\ref{potential}). We therefore neglect these terms for the remainder of the paper \footnote{The impact of the $Z_2$ breaking mass terms on the vacuum stability were discussed in a recent paper \cite{pko} where a large tree-level shift in the Higgs quartic coupling is possible even with a small mixing.}. We note that such terms can be forbidden in extensions of the minimal model, such as models with a global or local $U(1)$ symmetry. We will discuss such UV extensions of the model later.   

The Lagrangian (\ref{eq:Lferm}) yields the mass mixing matrices for the charged fermions,
\begin{equation}
{\cal M}'_e  =
\left( \begin{array}{cc}
M_l +x_l w/\sqrt{2} & \tilde y_e v /\sqrt{2} \\
 {y_e v}/{\sqrt{2}} & M_e + x_e w/\sqrt{2} 
\end{array}
 \right) , ~~~~  
 \end{equation}
and an analogous mass matrix in the neutral fermion sector. We bring the system to the physical basis through a bi-unitary transformation,
\begin{equation}
{\cal M}_e = {\cal U}_e^\dag {\cal M}_e' {\cal V}_e = {\rm diag}(m_{e_1}, m_{e_2}).
\end{equation}
For simplicity we will for the remainder of this paper specialize to the case 
\begin{equation}
M_l = M_e \equiv M,~~~ x_l = x_e \equiv x,~~~ y_e = \tilde y_e \equiv y.
\label{maximal}
\end{equation}
For a fixed lightest eigenvalue $m_{e_1}$ this choice will maximize the $h \rightarrow \gamma\gamma$ enhancement. 
The mass eigenvalues in this case are $m_{e_1 , e_2} = M+x w/\sqrt{2} \mp y v/\sqrt{2}$, and the mixing angle is maximal, $\theta_e = \pi/4$.

\paragraph{\bf Diphoton rate vs. vacuum stability}

\begin{figure}\centerline{
\includegraphics[width=0.45\textwidth]{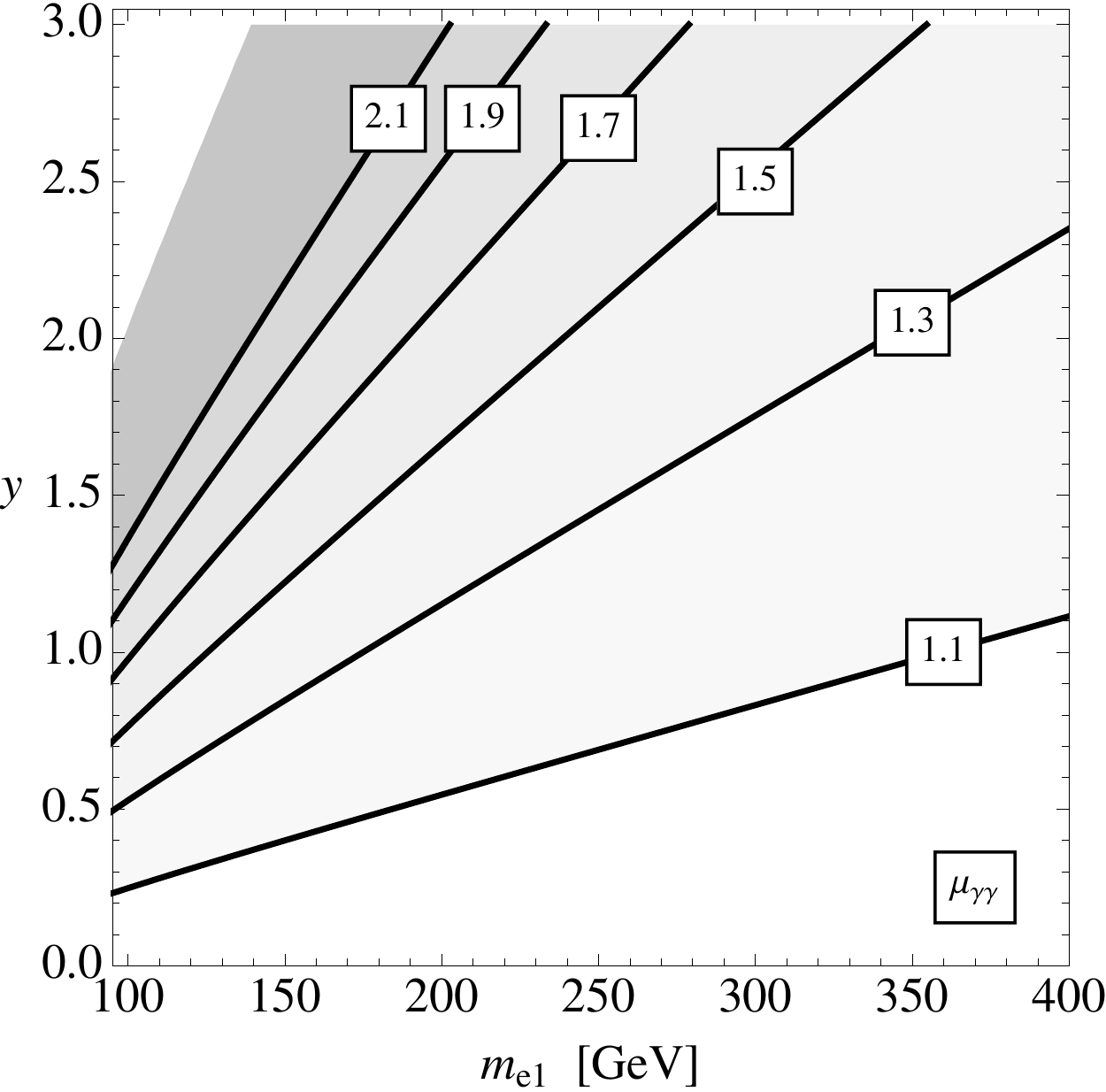}~~
}
\caption{\small Isocontours of the signal strength $\mu_{\gamma\gamma}$ in the $m_{e_1} - y$ plane. Here we have fixed vector-like masses via the relations (\ref{maximal}) and also taken $\theta = 0$ and $x=0$. }
\label{fig:rgy}
\end{figure}
The modification to the diphoton signal strength $\mu_{\gamma\gamma}$ in the model presented above is found to be
\begin{eqnarray}
\mu_{\gamma\gamma} & = &   \bigg\vert  c_\theta +  \frac{ v }{  \sqrt{2}   A_\gamma^{SM} } 
 \bigg[    y c_\theta
\left(  \frac{A_{1/2}(m_{e_1}) }{m_{e_1}} - \frac{A_{1/2}(m_{e_2}) }{m_{e_2}} \right)  \nonumber \\
&& ~~~+ ~ x  s_\theta  
\left(  \frac{A_{1/2}(m_{e_1}) }{m_{e_1}} +\frac{A_{1/2}(m_{e_2}) }{m_{e_2}} \right) 
   \bigg]  \bigg\vert^2,
  \label{rgam2} 
\end{eqnarray}
where $A_{1/2}$ is the standard loop function for fermions~\cite{Djouadi:2005gi}. We reiterate that this formula is valid in the case of common vector-like masses and Yukawa couplings displayed in Eq.~(\ref{maximal}). 

There are three effects at work in Eq.~(\ref{rgam2}). First, due to mixing with the singlet the Higgs has a reduced coupling to SM matter fields, and in particular the $W$-boson and top quark, which is captured by the first term in (\ref{rgam2}). Note that this also modifies the total Higgs production rate and total decay width by a factor of $\cos^2\theta$ which then cancel in the signal strength. Second, the Higgs couples directly to the charged vector-like leptons with Yukawa coupling $y$ which yield a 1-loop contribution to $h\rightarrow \gamma\gamma$. Finally, through mixing with the  singlet, the physical Higgs scalar inherits the direct couplings of $x$ of the singlet to charged fermions. 

We show in Fig.~\ref{fig:rgy} contours of the signal strength $\mu_{\gamma\gamma}$ in the $m_{e_1} - y$ plane, first  for the case of no mixing in the Higgs-singlet sector, $\theta = 0$, which is a good approximation when the singlet is heavy. The enhancement is maximized as the mass of the lightest charged fermion $m_{e_1}$ becomes smaller and the Yukawa coupling $y$ becomes larger, which can be easily understood by  examining the approximate formula (\ref{ssLET}) derived using the low energy theorem.

\begin{figure}\centerline{
\includegraphics[width=0.44\textwidth]{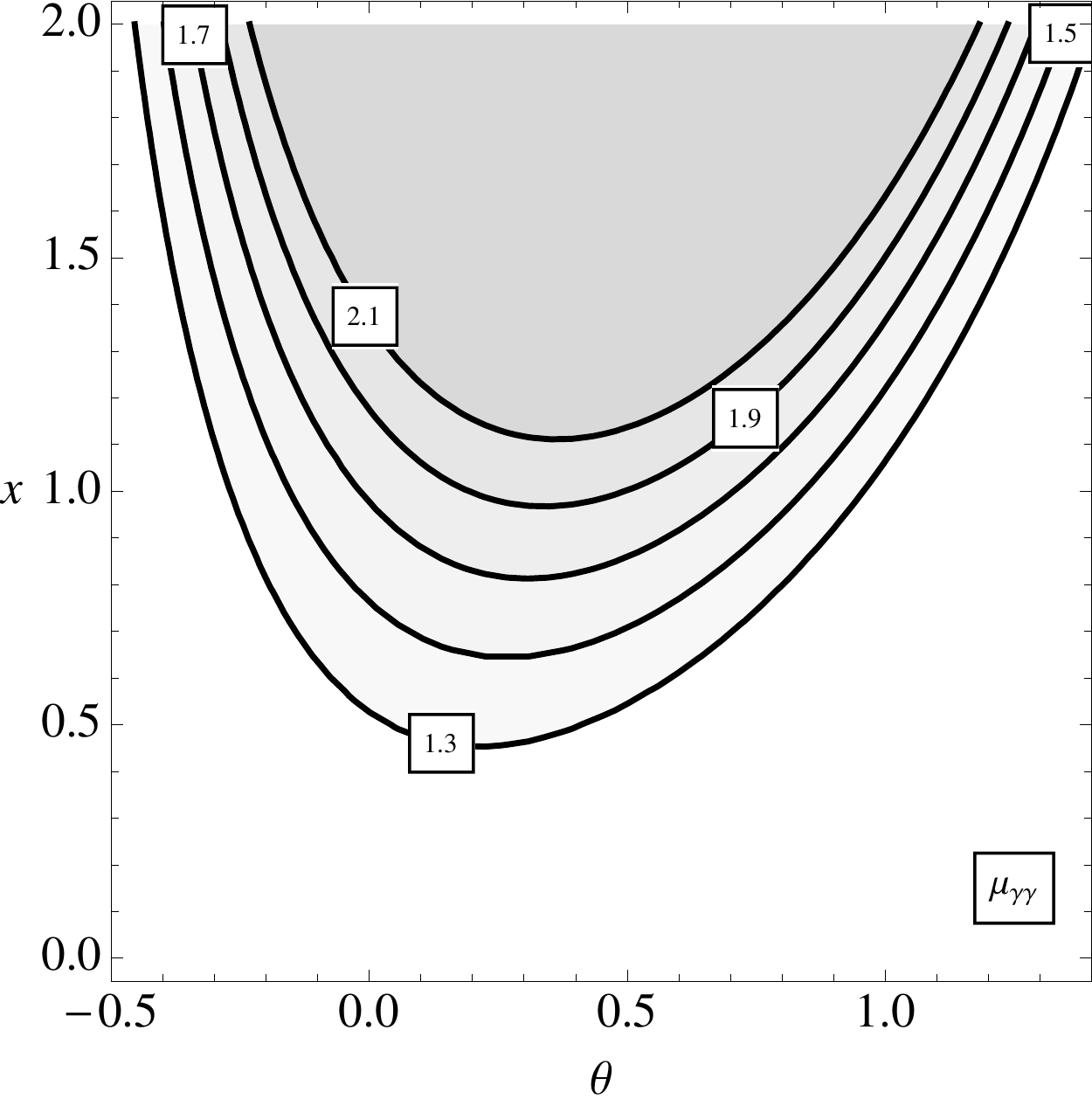}~~
}
\caption{\small Isocontours of the signal strength $\mu_{\gamma\gamma}$ in the $\theta - x$ plane. Here we have fixed the Higgs and singlet Yukawa couplings to be equal, $x=y$. We have also fixed the vector-like masses via the relations (\ref{maximal}) and taken $m_{e_1} = 100$ GeV.}
\label{fig:rgx}
\end{figure}

For lighter singlet masses close to the weak scale, achieving a sizable tree-level Higgs quartic coupling, or equivalently a large value of $\delta \lambda $ (defined in Eq.~(\ref{shift})) requires sizable Higgs-singlet mixing as is clearly shown in Fig.~\ref{fig:delThM}. In this case, a nonzero Yukawa coupling $x$ of the singlet to the vector-like leptons can have a substantial effect on the $h\rightarrow \gamma\gamma$ rate. We illustrate this in Fig.~\ref{fig:rgx} which displays contours of the the signal strength $\mu_{\gamma\gamma}$  in the  $\theta - x$  plane. Here we have chosen the Higgs and singlet Yukawa couplings to be equal, $x=y$. We have also fixed the lightest charged fermion mass eigenvalue to be $m_{e_1} = 100$ GeV. One can see that there exists an optimal mixing angle in terms of enhancing $\mu_{\gamma\gamma}$ with the smallest possible values of $x= y$. For example, to obtain an enhancement of $\mu_{\gamma\gamma}= 1.7$, the minimum possible Yukawa couplings occur at $x = y \approx 0.8$ which occurs for $\theta \approx 0.3$. 

\begin{figure}\centerline{
\includegraphics[width=0.45\textwidth]{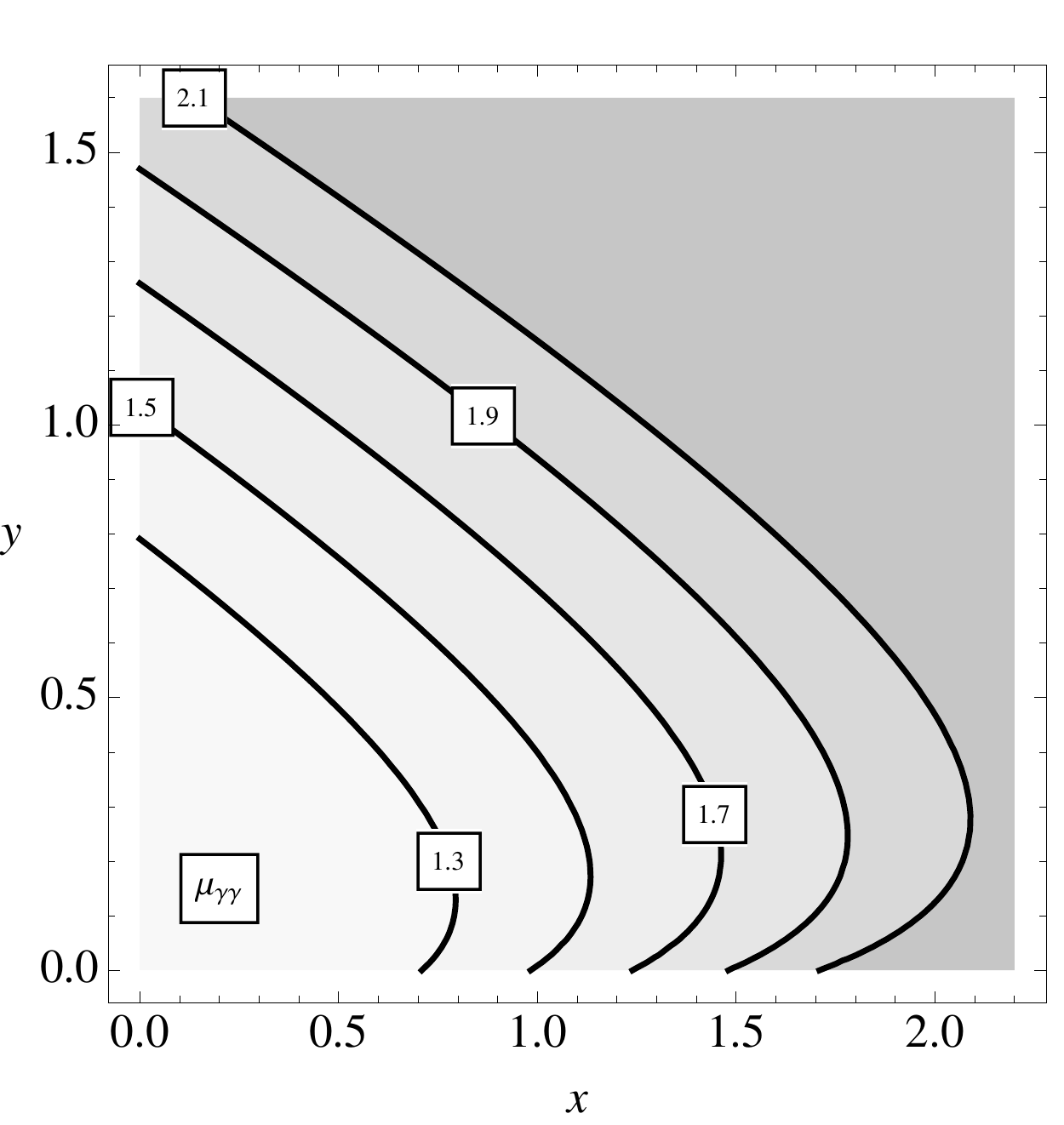}~~
}
\caption{\small Isocontours of the signal strength $\mu_{\gamma\gamma}$ in the $x - y$ plane. Here we have fixed vector-like masses via the relations (\ref{maximal}) with the lightest being $m_{e_1}=100$ GeV. We have also taken a mixing angle $\theta = 0.4$. }
\label{fig:rgxy}
\end{figure}

We also show in Fig.~\ref{fig:rgxy} values of $\mu_{\gamma\gamma}$ in the $x-y$ plane for a fixed mixing angle $\theta = 0.4$. For a fixed enhancement $\mu_{\gamma\gamma}$, increasing values of $x$ correspond to decreasing values of $y$, since for this value of the mixing angle the $x$ and $y$ contributions interfere constructively (see Eq.~(\ref{rgam})).

We now examine quantitatively the stability of the vacuum in this model. The RG equations for the model are presented in the Appendix. We will first consider a heavy singlet, in which case at the scale $Q = m_s$ one matches the full theory including the singlet to the SM + vector-like leptons as a low energy effective theory. Following this, we will investigate the possibility of a singlet near the weak scale, in which case a significant mixing between the Higgs and singlet is necessary to obtain a large Higgs quartic coupling.

\paragraph{Heavy singlet threshold}

We focus here on the case of a heavy scalar $m_s \gg v$. In this  regime, it is useful to consider the model in two stages. First  at energy scales $Q \ll m_s$ we integrate out the singlet and consider the effective theory, as in Eq.~(\ref{effpot}), which is simply the SM augmented with vector-like leptons. The effective Higgs quartic coupling, given in Eq.~(\ref{effQ}) runs quickly towards negative values due to the large Yukawa couplings of the new fermions. At the scale $m_s$ we match to the full theory, and follow the evolution of the couplings $\lambda_H$, $\lambda_S$, and $\lambda_{HS}$. Clearly, for this mechanism to work, the singlet threshold must be below the would-be vacuum instability scale.
Therefore, a fairly robust conclusion is that the singlet should be lighter than $\sim 10$ TeV if $\mu_{\gamma\gamma}$ is enhanced by $\sim50\%$ or more. 

\begin{figure}\centerline{
\includegraphics[width=0.45\textwidth]{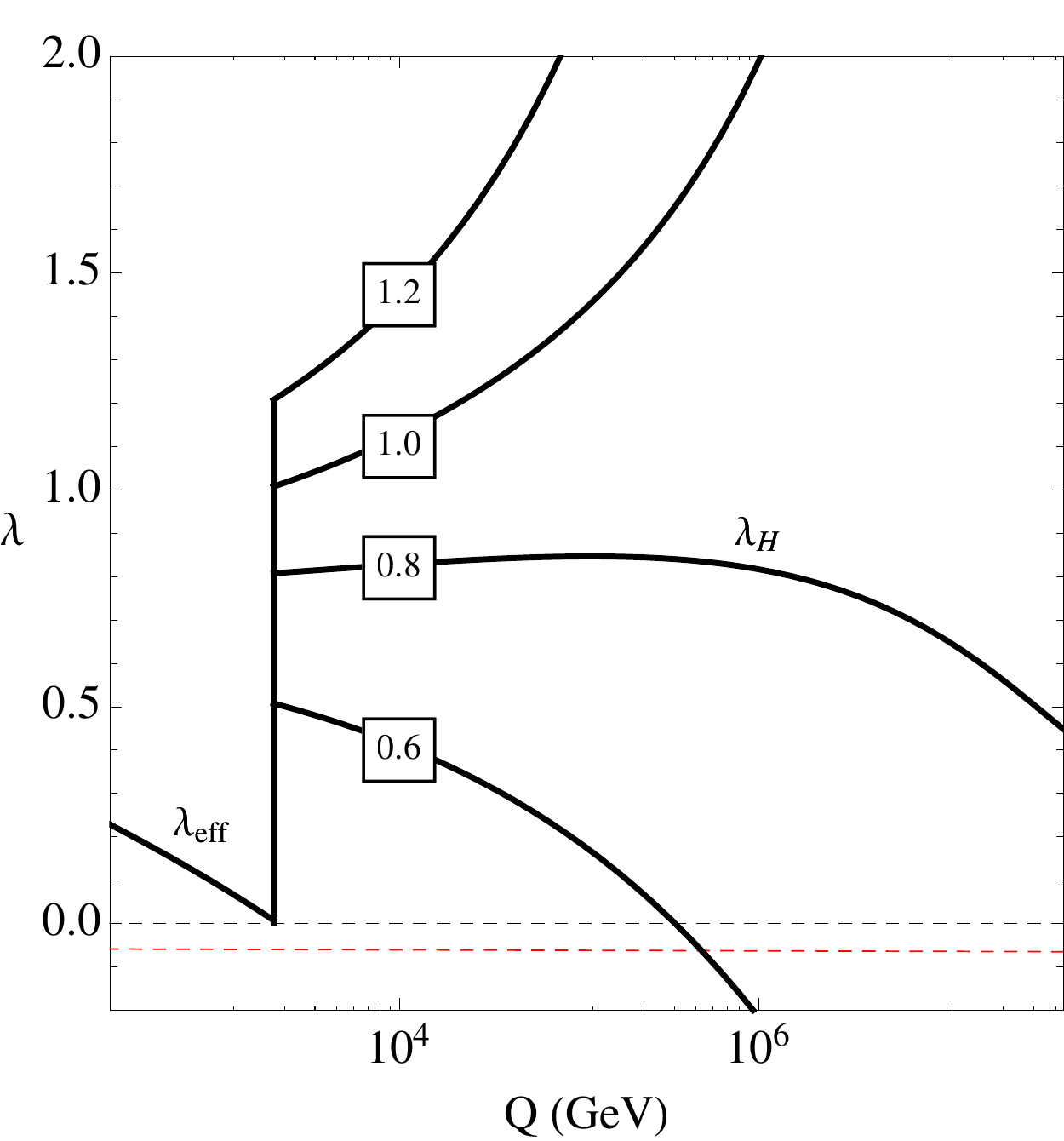}~~
}
\caption{\small Running of the Higgs quartic couplings for a heavy singlet $m_s = 2$ TeV: 1) effective quartic coupling $\lambda_{\rm eff}$ for $Q < m_s$, and 2) quartic coupling $\lambda_H$ for $Q >  m_s$. We have shown values of $\delta \lambda= 0.6,0.8,1.0,1.2$.
In this plot we have fixed $y = 1$, $x= 0$ at the weak scale.}
\label{fig:lambdaQ}
\end{figure}

We first show the effect of the singlet scalar threshold in Fig.~\ref{fig:lambdaQ}, which displays the evolution of the quartic coupling $\lambda_{\rm eff}$ defined in Eq.~(\ref{effQ}), valid for $Q < m_s$ and the full Higgs quartic coupling $\lambda_H$ for $Q > m_s$. For $Q>m_s$ we have displayed the evolution for several values of the tree-level shift $\delta \lambda$.  Here we have assumed a Yukawa coupling $y = 1$ at the weak scale, corresponding to a signal strength $\mu_{\gamma\gamma}\sim 1.7$ for $m_{e_1} \sim 100$ GeV.  
Clearly the addition of the singlet allows for a viable theory up to much higher scales. 
We see that that there are two possible outcomes depending on the value of $\delta\lambda$. For small values of $\delta\lambda$ the quartic coupling $\lambda_H$ eventually runs negative leading to a vacuum stability bound. For large values of $\delta\lambda$, on the other hand, the quartic coupling  $\lambda_H$ grows until it becomes nonperturbative. 

\begin{figure}\centerline{
\includegraphics[width=0.45\textwidth]{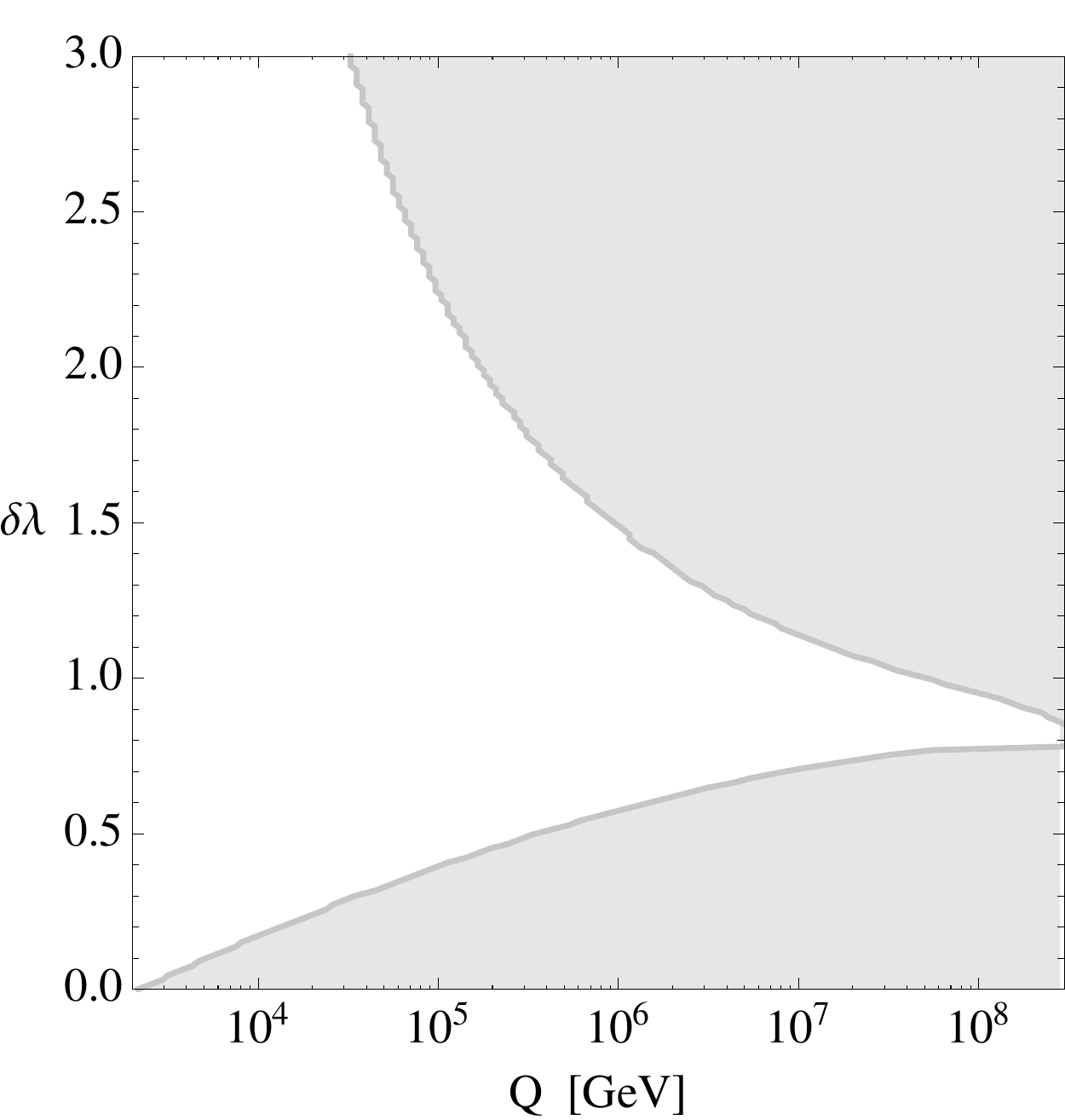}~~
}
\caption{\small Allowed values of $\delta\lambda$ as a function the energy scales $Q$. In this example we have taken $m_s=2$ TeV.
The lower shaded region represents values of $\delta\lambda$ for which the potential is not absolutely stable, 
while the upper shaded region indicates values of $\delta\lambda$ for which the Higgs quartic coupling  $\lambda_H > 4\pi$ is nonperturbative.
In this plot we have fixed $y = 1$ at the weak scale (corresponding to a signal strength enhancement $\mu_{\gamma\gamma}\sim 1.7$ for $m_{e_1} \sim 100$ GeV). 
The Higgs-singlet mixing is small in this regime and we have also assumed vanishing singlet Yukawa couplings $x=0$.
}
\label{fig:deltaQ}
\end{figure}

Furthermore, in Fig.~\ref{fig:deltaQ} we show the (absolute) vacuum stability bounds and the the perturbativity bound for the quartic coupling shift $\delta \lambda$ (defined at the scale $m_s$)  as a function of the RG scale $Q$. Again we have assumed a Yukawa coupling $y = 1$ at the weak scale, corresponding to a signal strength $\mu_{\gamma\gamma}\sim 1.7$ for $m_{e_1} \sim 100$ GeV. We see that in this case it is possible to have a viable theory up to scales $Q \sim 10^9$ GeV.

As emphasized in Ref.~\cite{EliasMiro:2012ay}, the absolute stability condition in the full theory at scales $Q \sim m_s$ is given by $\lambda_H  \lambda_S > \lambda^2_{HS}$, or equivalently $\lambda_H > \delta\lambda$. This condition need only be satisfied at scales near $m_s$, while for scales $Q \gg m_s$ the absolute stability condition quickly tends toward $\lambda_H >0$.

\paragraph{Singlet near the weak scale}

We have demonstrated that a heavy singlet $m_s \gg v$ can substantially raise the vacuum stability scale in the presence of vector-like leptons with ${\cal O}(1)$ Yukawa couplings to the Higgs. We now consider the situation in which the singlet is light, $m_s \sim v$. The new element in this case is the need for a large mixing angle $\theta$ between the Higgs and the singlet, which is required in order to obtain a large tree level quartic coupling $\lambda_H$ for $m_s \sim v$ (see Fig.~\ref{fig:delThM}). 

In the absence of direct couplings $x$ of the singlet to the vector-like leptons, the Yukawa couplings of the Higgs to the vector-like leptons $y$ must be increased as $\delta \lambda$ and, thus, the mixing angle are increased in order to obtain the desired enhancement to the signal strength $\mu_{\gamma\gamma}$. This is clear from examining Eq.~(\ref{rgam2}).  Such large Yukawa couplings cause the Higgs quartic coupling $\lambda_H$ to run even faster toward negative values. We have numerically investigated this case in detail, and the end result is that the required increase in the Yukawa couplings offsets, to a large degree, the benefit of the larger Higgs quartic coupling, so that there is still a vacuum stability issue to be addressed at low scales, $\Lambda \lesssim O(10~{\rm TeV})$

\begin{figure}\centerline{
\includegraphics[width=0.435\textwidth]{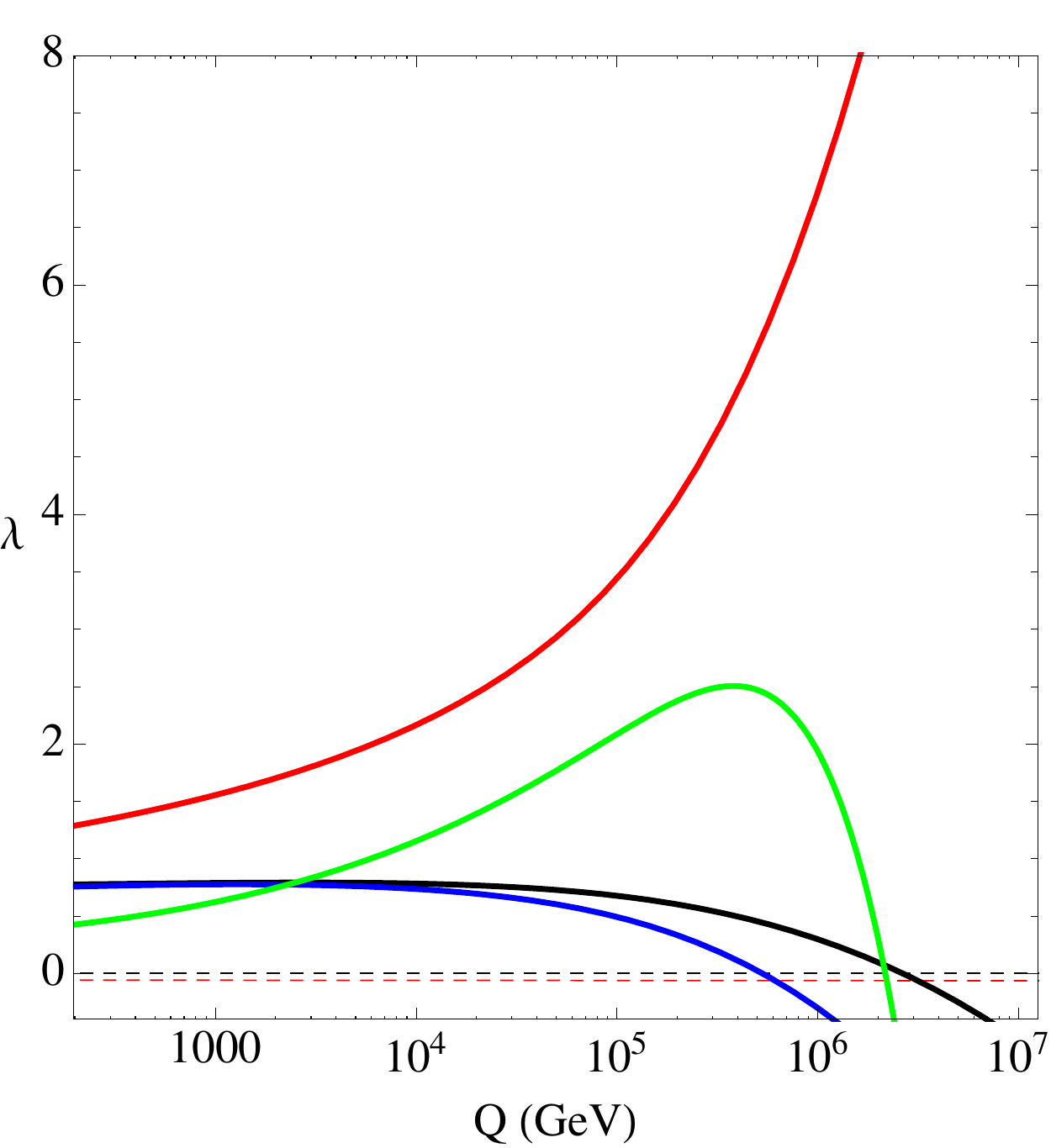}~~
}
\caption{\small Evolution of $\lambda_H$ (black), $\lambda_S$ (red), and $\lambda_{HS}$ (blue) as a function of the RG scale $Q$. 
The green line represents the quantity $\lambda_H \lambda_S - \lambda_{HS}^2$. The last quantity must be positive for vacuum stability when $\lambda_{HS}<0$.
Here we have fixed 
$m_s = 400$ GeV, $\lambda_H = 0.8$, $\lambda_{S} = 1.3$, $\lambda_{HS}=0.8$, $y = 1.0$, $x = 0.7$. The mixing angle in this case is  $\theta \simeq 0.5$.
This leads to an enhancement $\mu_{\gamma\gamma} = 1.7$ when the lightest charged vector-like lepton mass is 100 GeV.
}
\label{fig:LamX}
\end{figure}

Turning on Yukawa couplings $x$ of the singlet to the vector-like leptons can help in this respect.  Indeed, one can take advantage of the mixing angle: as can be seen from Fig.~\ref{fig:rgx}, by having both Higgs and singlet Yukawa couplings $y$ and $x$, there is an optimal mixing angle which minimizes the magnitude of such couplings for a fixed value of the enhancement of $\mu_{\gamma\gamma}$.  Minimizing the Higgs Yukawa coupling $y$ causes $\lambda_H$ to run more slowly, and the mixing angle allows for a large tree level coupling $\lambda_H$. 
However, at the same time, the singlet Yukawa couplings $x$ yield a negative contribution to the singlet quartic coupling $\lambda_S$ and cross coupling $\lambda_{HS}$ proportional to $-x^4$, so that one must now worry about these couplings becoming negative. In Fig.~\ref{fig:LamX} we show the evolution of $\lambda_H$, $\lambda_S$, and $\lambda_{HS}$ as a function of the RG scale $Q$. We see that the vacuum stability can be maintained to scales $Q ~\sim 10^6$ GeV  for appropriate choices of the parameters. The parameters chosen for this example point lead to an enhancement $\mu_{\gamma\gamma} \sim 1.7 $ for light vector-like fermions near the LEP bound. 

\paragraph{Summary}

We have demonstrated that the presence of a new singlet scalar field can alleviate a potential vacuum instability problem at low scales caused by new charged fermions responsible for enhancing the $h\rightarrow \gamma\gamma$ rate. This mechanism is very efficient for a heavy scalar, $m_s \gtrsim$ TeV, in which the theory can be perturbative and have a stable vacuum to very high scales, $\Lambda \sim 10^9$ GeV. For a light singlet near the weak scale, $m_s \sim v$, a large mixing angle between the Higgs and singlet is necessary in order to obtain a large ${\cal O}(1)$ value for the Higgs quartic coupling. In this case, one can take advantage of such a large mixing by coupling the singlet to the new charged fermions, so that the Higgs inherits these couplings through mixing. In this case as well, the theory can be viable to  a very high scale, $\Lambda \sim 10^6$ GeV. This case is more interesting phenomenologically since the light singlet can be produced at the LHC. We now move to an investigation of the singlet phenomenology. 

\section{ Singlet Phenomenology}

We now explore several phenomenological consequences of our scenario, including the effects of the vector-like leptons and singlet scalar in the Higgs signal strength data, the interplay between the precision electroweak data and the stability of the vacuum, and the signatures of the singlet at the LHC.

\paragraph{\bf Higgs-singlet mixing vs. signal strength data}

For singlets near the weak scale, a large tree level shift $\delta\lambda$ to the Higgs quartic coupling requires sizable mixing between the Higgs and the singlet. Such a mixing will affect the Higgs couplings and the predictions for the signal strength in the various channels $\mu_i$. It is thus of interest to determine the range of mixing angles which provide a good description of the signal strength data. 

\begin{figure}\centerline{
\includegraphics[width=0.45\textwidth]{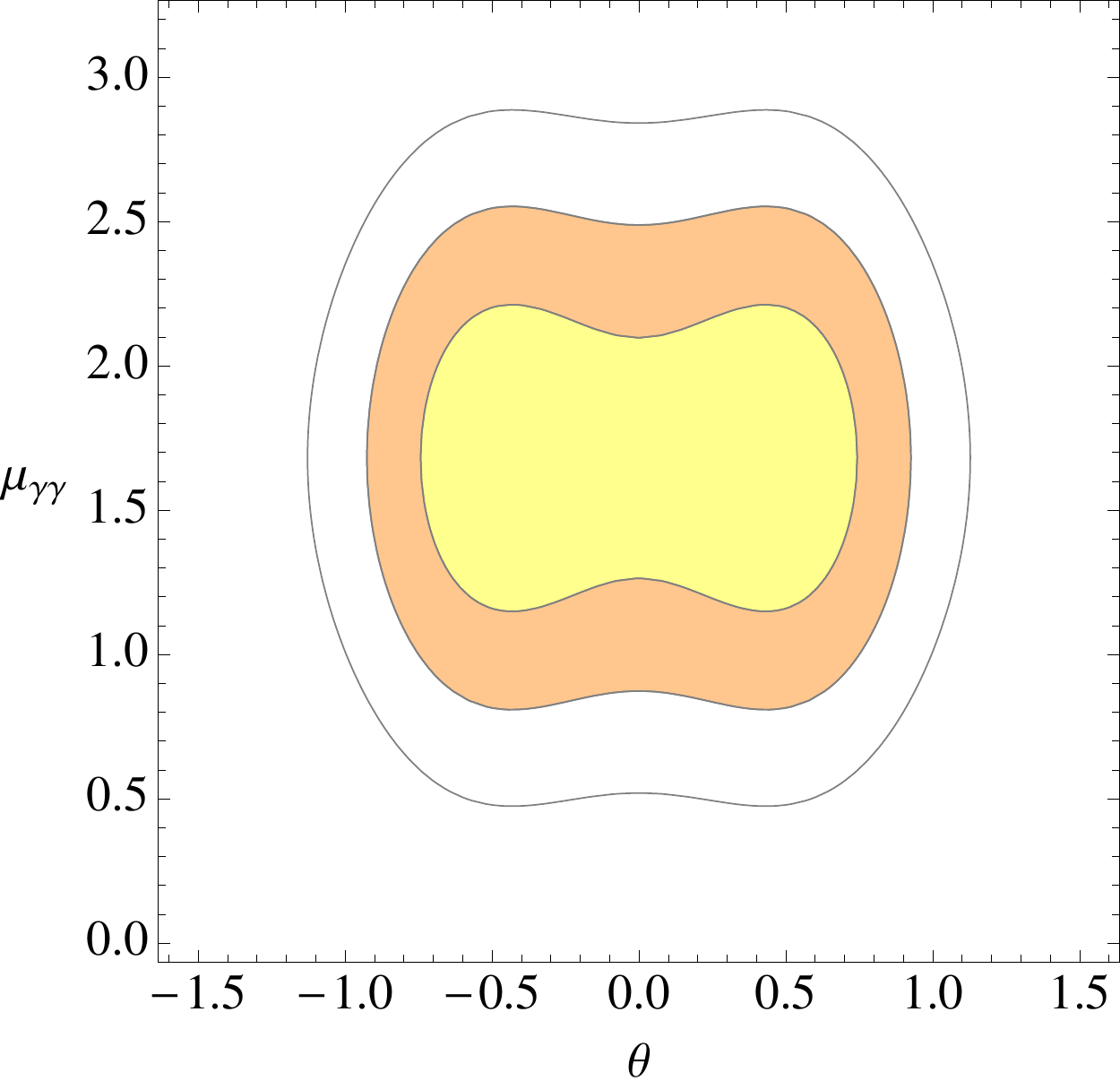}~~
}
\caption{\small  The $1,2,3\sigma$ CL regions in $\theta-\mu_{\gamma\gamma}$ plane from a fit to the Higgs signal strength data. }
\label{fig:higgs}
\end{figure}
The mixing between the singlet and the Higgs alters the signal strength prediction as $\mu_{i} = \cos^2\theta$ for $i = WW, ZZ, b\bar b, \tau \bar \tau$. The exception is the
 $\gamma\gamma$ channel, which is modified according to Eq.~(\ref{rgam2}). To get a feeling for the allowed values of the mixing angle, we have performed a fit to the signal strength data in the $\gamma\gamma, WW, ZZ, b\bar b, \tau \bar \tau$ data from the ATLAS~\cite{ATLAS-Higgs}, CMS~\cite{CMS-Higgs}, and Tevatron~\cite{:2012zzl} experiments. Since several parameters enter into the prediction for the diphoton signal strength (explicitly, $m_{e_1,e_2}$, $y$, $x$, $\theta$), we find it  most illuminating simply treat 
$\mu_{ \gamma \gamma } $ as a free parameter and perform a two parameter fit allowing $\theta$ and $\mu_{\gamma\gamma}$ to vary freely. 

We now summarize the results of the fit: The best-fit point is  
$(\theta, \mu_{\gamma\gamma})  = (0.4, 1.7)$ with a goodness-of-fit described by $\chi^2/{\rm d.o.f} = 8.1/11$.
This is to be compared with the SM which gives a goodness-of-fit of  $\chi^2/{\rm d.o.f} = 12.8/13$. That a small mixing angle is preferred is simply due to the fact that this leads to a slight suppression, $\mu_{VV} \sim 0.8$  for $VV = WW, ZZ$ channels, as shown by the pattern in Eq.~(\ref{pattern}).
In Fig.~\ref{fig:higgs} we show the allowed regions in the $\theta-\mu_{ \gamma \gamma }$ plane. We observe that a sizable mixing angle is still allowed by the data, up to values of $\theta\sim 0.7$. See also Refs.~\cite{singlets} for recent studies of the effects of singlet-Higgs mixing on the Higgs data.

\paragraph{\bf Electroweak precision tests}

\begin{figure}\centerline{
\includegraphics[width=0.47\textwidth]{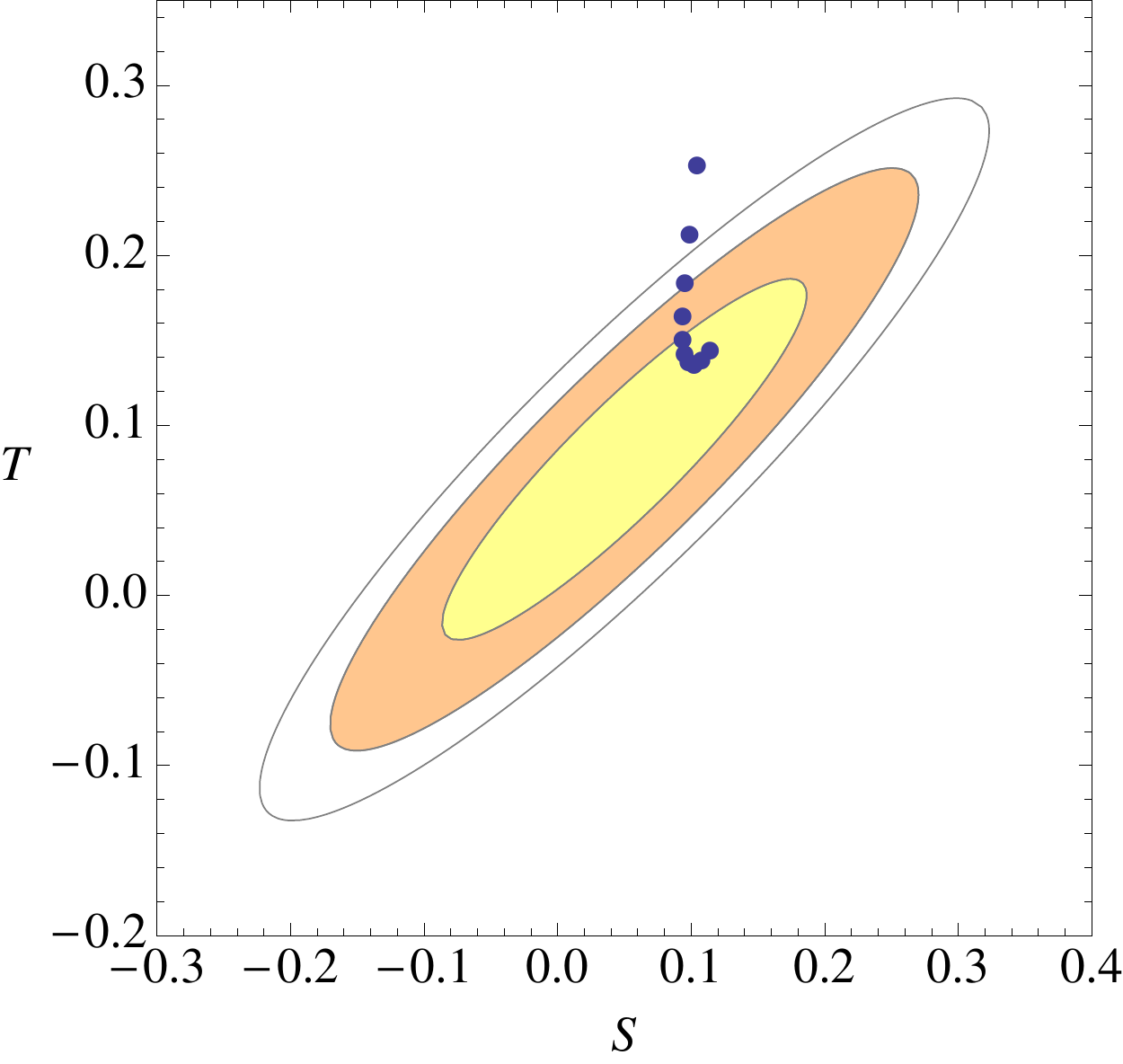}~~
}
\caption{\small Oblique parameters $S$ and $T$. Here we show the 1,2,3$\sigma$ CL regions in the $S-T$ plane. The blue points show the effect of changing the mixing angle from $\theta = 0$ (where the model is excluded at more than 3$\sigma$ CL) to $\theta = 0.5$ (where the model lies within the 1$\sigma$ CL region.
For these points we have fixed $y = x$ to the values along the $\mu_{\gamma\gamma} = 1.7$ contour in Fig.~\ref{fig:rgx}. We have also fixed $m_{e_1}= 100$ GeV, $ y_\nu = x_\nu = 0$, and $m_s = 500$ GeV.}
\label{fig:ST}
\end{figure}
The new vector-like leptons give contributions to the oblique parameters $S$ and $T$~\cite{Peskin:1991sw}. 
In particular, as emphasized in Refs.~\cite{Joglekar:2012vc,ArkaniHamed:2012kq}, there is a trade off between minimizing the $T$ parameter and obtaining a high vacuum instability scale. In order to minimize the $T$ parameter, one would like to impose an approximate custodial symmetry in the Lagrangian (\ref{eq:Lferm}), with $M_e \sim M_\nu$ and $y_e \sim y_\nu$. However, additional large Yukawa couplings $y_\nu$ will drive the Higgs quartic couplings towards negative values at a faster rate, so that the instability scale is lower than if those couplings were small. 

In the model with a singlet scalar there are new effects due to Higgs - singlet mixing and the Yukawa couplings $x$ of the singlet to the vector-like leptons. If the singlet is heavy $m_s \gg m_h$, a non-zero mixing angle $\theta$ between the Higgs and the singlet mimics the effect of a heavy SM Higgs boson, causing a positive shift in the $S$ parameter and a negative shift in the $T$ parameter. Furthermore, as shown in Fig.~\ref{fig:rgx}  nonzero singlet Yukawa couplings $x$ allow one to utilize the Higgs singlet mixing to enhance the $h\rightarrow \gamma\gamma$ rate with somewhat smaller Higgs Yukawa couplings $y$, which reduces the $T$ parameter. 

These two effects allow one to alleviate the constraints from electroweak precision data provided the Higgs-singlet mixing angle is not too small, suggesting a singlet not far above the weak scale.
We illustrate this in Fig.~\ref{fig:ST} which displays several model parameter points in blue in overlaid in the $S-T$ plane. For the $S-T$ allowed regions we use the recent updates from the Gfitter group~\cite{Baak:2012kk}. In these model points, we have fixed the diphoton enhancement to a value $\mu_{\gamma\gamma} = 1.7$. We vary the mixing angle from $\theta = 0$, where the model is excluded at more than 3$\sigma$ confidence, to $\theta = 0.5$ at which point the model lies within the 1$\sigma$ $S-T$ ellipse. To obtain the enhancement $\mu_{\gamma\gamma} = 1.7$ we have 
taken common Yukawa couplings $y=x$ fixed to values along the $\mu_{\gamma\gamma} = 1.7$ contour in Fig.~\ref{fig:rgx} as the mixing angle varies. We have furthermore fixed  $m_{e_1}=100$ GeV and $m_s = 500$ GeV.

\paragraph{\bf Signatures of the singlet at LHC}  

Through mixing between the singlet and Higgs, the physical singlet scalar $s$ inherits the couplings of the Higgs to SM fields, which are rescaled by a universal mixing factor $\sin \theta$. Singlets can then be directly produced at LHC and subsequently decay to SM particles. The phenomenology of singlet scalars has been thoroughly investigated in the literature~\cite{singletpheno}. In our model, singlets can also decay to vector-like lepton pairs as well as light Higgs pairs if kinematically allowed. These decay modes can not only provide new discovery channels, but also relax possible bounds coming from the null results of Higgs searches for singlet masses $m_s \leq 600$~GeV.

The singlet partial decay widths into SM particles, vector-like lepton pairs, and $hh$ are, respectively,
\begin{eqnarray}
\Gamma(s \to {\rm SM}) &  = & \Gamma(h \to {\rm SM})_{{\rm SM}} \,\sin^2 \theta, \\
\Gamma(s \to e_1\bar  e_1) &  = & \frac{g_{s e_1 e_1}^2 m_s}{8\pi} \, \left( 1- \frac{4 m_{e_1}^2}{m_s^2} \right)^{3/2}, \\
\Gamma(s \to hh) &  = & \frac{g_{shh}^2}{32\pi m_s} \, \sqrt{1-\frac{4 m_h^2}{m_s^2} },
\end{eqnarray}
where a subscript SM implies the width calculated in pure SM evaluated at the mass $m_s$ of the singlet, and $e_1$ is the lightest charged fermion mass eigenstate. When needed throughout this section, we assume maximal charged fermion mixing (as in Eq.~(\ref{maximal})) to maximally enhance the diphoton rate, a lightest charged fermion mass $m_{e_1}=100$ GeV, and adjust the Yukawa couplings $x=y$ to achieve $\mu_{\gamma \gamma}=1.7$ as we vary Higgs mixing $\theta$, as in Fig.~\ref{fig:rgx}. With maximal fermion mixing, the physical couplings of $s$ are given by
\bea
g_{se_1 e_1} &=& \frac{1}{\sqrt{2}} (-y \sin \theta  + x \cos \theta ),
\label{eq:se1e1} \\
g_{shh} &=& \cos^3 \theta \lambda_{HS} w + \cos^2 \theta \sin \theta( 3 \lambda_H - 2 \lambda_{HS} ) v \nonumber\\
& + & \cos \theta \sin^2\theta(3 \lambda_S - 2 \lambda_{HS})w + \sin^3 \theta \lambda_{HS} v.~~~~~~~
\eea

\begin{figure}\centerline{
\includegraphics[width=0.47\textwidth]{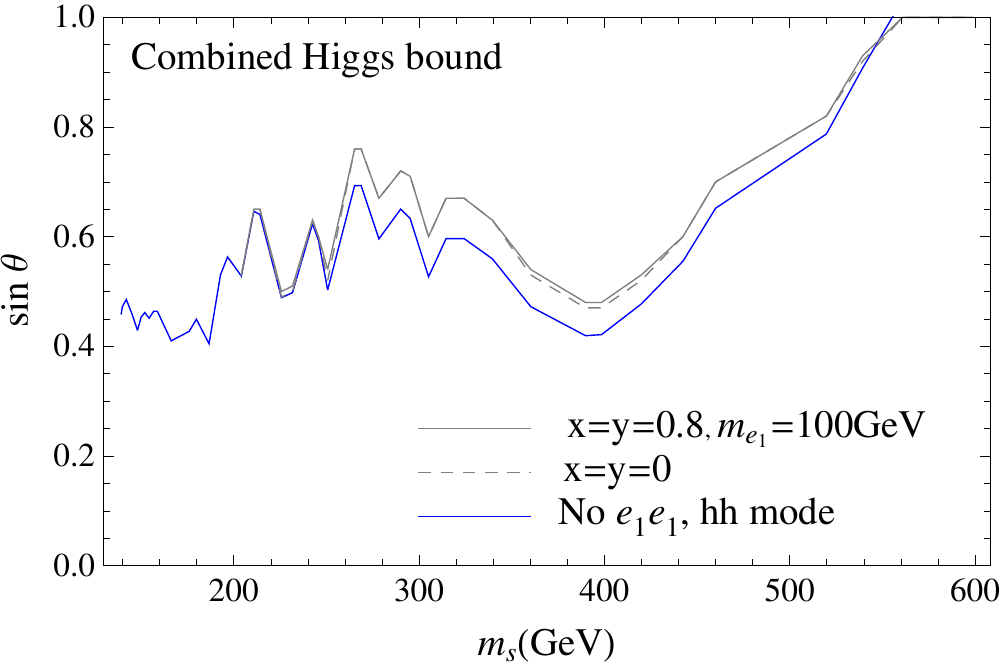}~~
}
\caption{Upper bound on Higgs-singlet mixing from all Higgs searches at ATLAS. For each line shown, we allow singlets to decay to different set of final states: pure SM(blue solid), additionally to $hh$ (gray dashed), additionally to 100 GeV lepton pair (gray solid). $x=y=0.8$ is used when necessary as this choice approximately achieves $\mu_{\gamma\gamma} \sim 1.7$ with $\theta$ around upper bound.}
\label{fig:bound_mixing}
\end{figure}
ATLAS and CMS have presented bounds on the total signal strength modifier $\mu$ for Higgs masses
up to $m_h=600$ GeV, which can be reinterpreted as a bound on the singlet mass - mixing angle parameter space. 
The prediction for the signal strength in our model is
\begin{equation}
\mu \equiv \frac{ \sigma(gg \to s) \! \cdot \!  {\rm Br}(s \to {\rm SM}) }{ \sigma(gg \to h)_{ {\rm SM}}  \! \cdot \!  {\rm Br}(h \to  {\rm SM})_{ {\rm SM}} } = \sin^2 \theta \,  {\rm Br}(s \to  {\rm SM}).
\end{equation}
Using the ATLAS combined results \cite{ATLAS-Higgs}, we obtain a bound on the mixing angle in Fig.~\ref{fig:bound_mixing} as a function of the mass of the singlet. For $m_s \lesssim 500$ GeV, this bound is more constraining than the one obtained from the best-fit of the 125 GeV resonance in
Fig.~\ref{fig:higgs}, although one can still have a fairly large mixing of $\sin \theta \lesssim 0.5$. 
The additional decay modes of the singlet into a Higgs pair and vector-like lepton pair have only a moderate effect in relaxing the limits because the upper bound corresponds to a sizable mixing angle and thus a significant singlet decay width through Higgs mixing. 
The $hh$ decay mode has a larger effect than the $e_1 \bar e_1$ mode because the physical coupling of $s$ to the vector-like leptons in Eq. (\ref{eq:se1e1}) is small due to a partial cancellation between the $y$ and $x$ contributions when the fermion mixing and Higgs-singlet mixing is large. 
However, for smaller Higgs mixing $\sin \theta \lesssim 0.4$, this coupling becomes larger and branching ratio into $e_1 \bar  e_1$ mode dominates. 

\begin{figure}\centerline{
\includegraphics[width=0.5\textwidth]{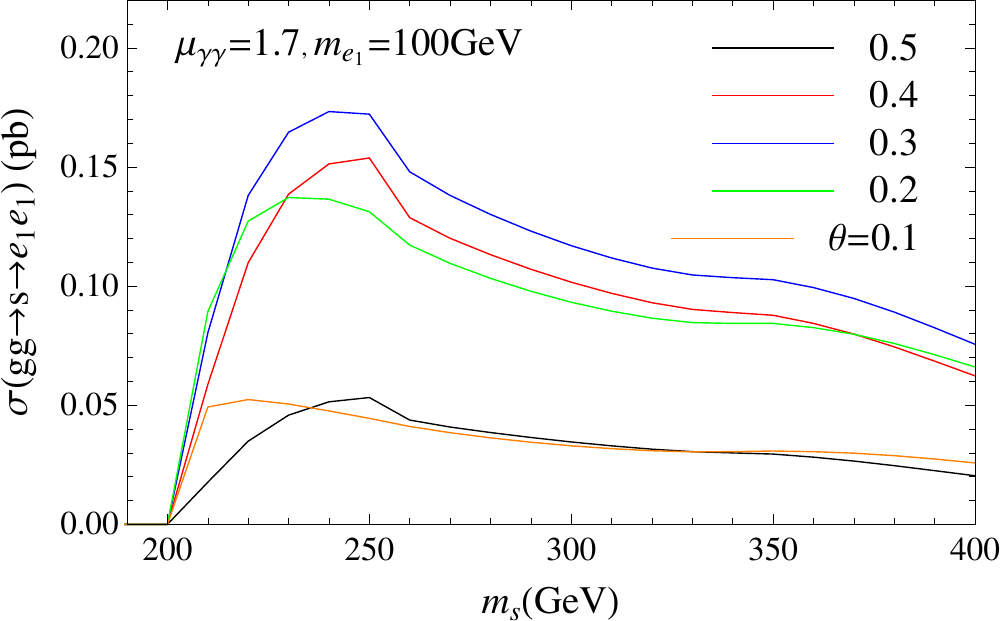}~~
}
\caption{$\sigma(gg\to s) \cdot {\rm Br}(s \to e_1 \bar e_1)$ in units of pb is plotted. Higgs mixing is varied within the allowed region. $m_{e_1}=100$GeV is assumed and Yukawa couplings $x=y$ are chosen to achieve $\mu_{\gamma \gamma} = 1.7$ for each choice $\theta$, as in Fig.~\ref{fig:rgx}. }
\label{fig:lepton_prod}
\end{figure}

Vector-like lepton pairs can be produced at the LHC through resonant production of $s$ via gluon fusion followed by its subsequent decay. In the limit of negligible Higgs-singlet mixing, vector-like lepton pairs are produced only through the Drell-Yan process. For instance, in the case of maximal mixing the cross section for production of $m_{e_1}=100$ GeV leptons is $0.4$ pb. The additional production of vector-like lepton pairs in our model is calculated in Fig.~\ref{fig:lepton_prod}. The rate of vector-like lepton pair production through $s$ decay can be as large as $40\%$ of Drell-Yan production. 
In the region of small mixing, vector-like lepton pair production increases as the mixing angle increases, due to the rising gluon fusion rate. However, the vector-like lepton production rate reaches a maximum at $\theta \sim 0.3$ and decreases back to zero because the branching ratio for $s\rightarrow e_1 \bar e_1$ is suppressed by due to small $g_{se_1 e_1}$ coupling and the increasing decay width of the singlet to SM modes via Higgs portal mixing. We note that the mixing $\theta \sim 0.3$ giving the largest production also provides a good fit to the 125 GeV Higgs signal strength data, as shown in Fig.~\ref{fig:higgs} and is consistent with constraints from precision electroweak data discussed above.

Thus, if the production rate of vector-like leptons is observed to be higher than that expected from Drell-Yan processes alone, it might suggest the existence of resonantly produced singlet scalar bosons. 
Another important test for the existence of $s$ at the LHC will be searches for Higgs-like states extending to higher masses $m_s > 600$ GeV using larger datasets.
A future high energy lepton collider would also allow detailed tests of this scenario, through extended Higgs resonance searches at high mass and measurements of the angular correlation of final state SM leptons (to which our vector-like leptons are assumed to decay) 
which can test the spin of $s$-channel mediator, $s$ and/or $\gamma$.

\section{UV extensions of the minimal model}

While we have investigated the minimal scenario that includes a real scalar singlet, it is natural to consider the singlet and vector-like leptons as part of a new sector charged under a global or local $U(1)'$ symmetry.
In this case, the masses of vector-like leptons can be obtained from the singlet Yukawa couplings as follows,
\be
{\cal L}_{S,\,{\rm Yuk}} =- \lambda_l S l_4 {\tilde l}_4 - \lambda_e S^* e^c_4 {\tilde e}^c_4+{\rm h.c.},
\ee
where we have taken the $U(1)'$ charges of vector-like leptons to be $q(l_4)=-q(e^c_4)$ and $q({\tilde l}_4)=-q({\tilde e}^c_4)$ in order to allow the Higgs Yukawa couplings with vector-like leptons.
Then, after the singlet gets a VEV $v_s$, we obtain the vector-like masses as $m_l=\frac{1}{\sqrt{2}}\lambda_l v_s$ and $m_e=\frac{1}{\sqrt{2}}\lambda_e v_s$.

\paragraph{\bf Global $U(1)'$ PQ symmetry}

First, when the $U(1)'$ is a global PQ symmetry, there would be a massless Goldstone boson after the $U(1)'$ is broken spontaneously. In order to avoid a dangerous weak-scale axion, we need to break the $U(1)'$ explicitly. So, a $U(1)'$ breaking mass term can be introduced as $\frac{1}{2}m^{\prime 2}_{S}  S^2+{\rm c.c.}$, which can be comparable to the $U(1)'$-invariant singlet mass term. Then, the axion should be included in the low energy spectrum.  For instance, the $U(1)'$ soft breaking mass term can be given by a high-scale PQ symmetry breaking \cite{lpp1}, from $\frac{1}{M^2_P}\,\Phi^4 S^2+{\rm h.c.}$ where $\langle\Phi\rangle\sim 10^{10}\,{\rm GeV}$ contains the invisible axion for solving the strong CP problem. If the axion partner of the singlet scalar has weak-scale mass, it affects the RG evolution of the Higgs quartic coupling and may mix with the CP-even scalars if the $U(1)'$ breaking mass violates CP.

\paragraph{\bf Local $U(1)'$ gauge symmetry }

Second, when the $U(1)'$ is a local symmetry, the would-be Goldstone boson is eaten by a massive $Z'$ gauge boson. So, in the decoupling limit of the $Z'$ gauge boson, we only have to take into account the CP-even singlet scalar in the low energy spectrum in addition to the vector-like leptons. In this case, we can still obtain a tree-level shift in the Higgs quartic coupling from the mixing between the CP-even singlet scalar and the Higgs boson. Of course, if $Z'$ gauge boson has a weak scale mass, it should be included in the discussion as well. 
However, in the case of a local $U(1)'$ symmetry, we need to consider the anomaly cancellation conditions such as $[U(1)']^3, U(1)'-SU(2)^2_L, U(1)'-U(1)_Y^2$, and $[U(1)']^2-U(1)_Y$ anomalies. The anomalies coming from the vector-like leptons are, respectively,
\bea
A'_1&=&2\,q(l_4)^3+ 2\,q({\tilde l}_4)^3+q(e^c_4)^3+q({\tilde e}^c_4)^3,\\
A_2 &=&  \frac{1}{2}(q(l_4)+q({\tilde l}_4)), \\
A_1&=&  \frac{1}{2} (q(l_4)+q({\tilde l}_4))+q(e^c_4)+q({\tilde e}^c_4), \\
A^{\prime\prime}_1&=& -\frac{1}{2}(q(l_4)^2-q({\tilde l}_4)^2)+q(e^c_4)^2-q({\tilde e}^c_4)^2.
\eea
Then, for $q(l_4)=-q(e^c_4)$ and $q({\tilde l}_4)=-q({\tilde e}^c_4)$, we get
\bea
A'_1&=&q(l_4)^3+q({\tilde l}_4)^3,\quad  A_2=\frac{1}{2}(q(l_4)+q({\tilde l}_4)), \nonumber \\
A_1&=&-\frac{1}{2}(q(l_4)+q({\tilde l}_4)), \quad  A^{\prime\prime}_1=\frac{1}{2}(q(l_4)^2-q({\tilde l}_4)^2).~~~~~~
\eea
If $q(l_4)=-q({\tilde l}_4)$, we can cancel all the anomalies but the tree-level vector-like masses are allowed. If $q(l_4)\neq q({\tilde l}_4)$, we can forbid the tree-level vector-like masses but the anomalies cannot be cancelled simultaneously.
For instance, for $q(l_4)=q({\tilde l}_4)=1$, only $A^{\prime\prime}_1$ vanishes but the other anomalies are nonzero: $A'_1=2$, $A_2=1$ and $A_1=-1$.
 Therefore, we need extra vector-like leptons to cancel the anomalies. 
 For instance, another set of vector-like doublets and singlets with $U(1)'$ charges, $q(l_5)=q({\tilde l}_5)=-1$ and $q(e^c_5)=q({\tilde e}^c_5)=1$, can cancel the anomalies. In this case, we need an additional complex scalar to give them masses.  Since the extra vector-like fermions can also couple to the Higgs, they could affect the Higgs diphoton rate as well as the vacuum stability.

\paragraph{\bf Two Higgs doublet model }

 For more general $U(1)'$ charges of vector-like leptons, we need to extend the Higgs sector with two Higgs doublets for obtaining the Yukawa couplings to the SM quarks and leptons. Then, the masses of vector-like leptons can be generated by 
\be
{\cal L}_{{\rm Yuk}} =- \lambda_l S l_4 {\tilde l}_4 - \lambda_e S e^c_4 {\tilde e}^c_4-y_e H_d l_4 e^c_4-{\tilde y}_e H_u {\tilde l}_4 {\tilde e}^c_4  +{\rm h.c.}.
\ee
Since the trace of PQ charges of vector-like leptons is nonzero, electromagnetic anomalies do not vanish. Nonzero $U(1)'$ anomalies imply that the $U(1)'$ symmetry is necessarily anomalous so it should be a global symmetry \footnote{If the $U(1)'$ anomalies are cancelled by a Green-Schwarz mechanism, the $U(1)'$ symmetry can be of local symmetry origin.}. 
When the pseudo-scalar of the singlet couples to a singlet fermion dark matter, it can induce dark matter annihilation into monochromatic photons by anomaly interactions for the Fermi gamma-ray line at $130\,{\rm GeV}$  \cite{lpp2}. 
The CP-even scalar of the singlet still can improve the vacuum instability by the quartic coupling with the Higgs bosons. In the case where the extra Higgs states are decoupled, our previous discussion with the singlet scalar and the SM Higgs boson holds in the effective theory. Furthermore, new effects arise in two Higgs doublet models that can affect the Higgs couplings; for recent studies see Refs.~\cite{2HDM}.

\section{Conclusions}

Simple models utilizing new charged vector-like `leptons' or other color-singlet, electrically charged fermions to enhance the $h\rightarrow \gamma\gamma$ partial decay width require ${\cal O}(1)$ Yukawa couplings, which yield negative contributions to the $\beta$ function of the Higgs quartic coupling. The Higgs quartic coupling runs very quickly towards negative values, suggesting a dangerous vacuum instability issue and necessitating new UV physics at relatively low scales. We have investigated an economical mechanism in which the scalar sector is extended to include a singlet which develops a VEV and mixes with the Higgs. This allows for a much larger Higgs quartic coupling compared to the SM and thus a stable vacuum to much higher scales.

The new scalar singlet in this scenario must be relatively light in order for the mechanism to work, {\it e.g.} below a few TeV a large enhancement $\mu_{\gamma\gamma} \sim 1.7$.
Furthermore, if the singlet is close to the weak scale, obtaining a large Higgs quartic coupling necessarily requires sizable mixing between the singlet and the Higgs. Such a mixing can be utilized to further enhance the diphoton rate through Yukawa couplings of the singlet to the vector-like leptons, and can allow for good descriptions of the Higgs signal strength and precision electroweak data. Furthermore due to the mixing in the scalar sector, the physical singlet scalar picks up couplings to SM particles and thus can be produced at colliders. The singlet may be searched for in standard Higgs-like decay modes as well as through anomalous production of the new vector-like leptons.

\subsubsection*{Acknowledgements}

B.B. is supported by the NSF under grant PHY-0756966 and the DOE Early Career Award under grant
DE-SC0003930.

\newpage

\appendix*
\begin{widetext}
\def\theequation{\thesection A.\arabic{equation}}
\setcounter{equation}{0}
\section{Appendix~~~$\beta$ functions}
\label{ap:beta}

Here we present the $\beta$ functions for an extension of the SM containing vector-like leptons and a singlet scalar, which can be obtained from the general results of Refs.~\cite{RG}:

\begin{align}
16 \pi^2 \beta_{\lambda_H} &  =  12 \lambda_H^2 + \lambda_{HS}^2 +
 \lambda_H \left[ 12 y_t^2+ 4(y_e^2+\tilde y_e^2+ y_{\nu}^2 + \tilde y_{\nu}^2 )  - 9 g^2 - 3 g'^2   \right] \nonumber \\ &
   -12 y_t^4- 4(y_e^4+\tilde y_e^4+ y_{\nu}^4 + \tilde y_{\nu}^4 ) 
  + \frac{9 g^4}{4}+\frac{3 g'^4}{4}+ \frac{3 g^2 g'^2}{2}  \nonumber  \\
16 \pi^2 \beta_{\lambda_S} &  =  9 \lambda_S^2   +4 \lambda_{HS}^2   + 4 \lambda_S (2 x_l^2 +  x_e^2  +  x_\nu^2) 
-4 ( 2 x_l^4  +  x_e^4  +  x_\nu^4 ) 
~~\nonumber\\
16 \pi^2 \beta_{\lambda_{HS}} &  =  4 \lambda_{HS}^2  + 
 \lambda_{HS} \bigg[6 \lambda_H + 3 \lambda_S + 6 y_t^2+ 2(y_e^2+\tilde y_e^2+ y_{\nu}^2 + \tilde y_{\nu}^2 )  
  +2( 2 x_l^2+x_e^2+ x_{\nu}^2  )   - \frac{9 g^2}{2} - \frac{3 g'^2}{2}   \bigg] 
  \nonumber  \\  &
  -4 x_l^2 (y_e^2+\tilde y_e^2+ y_{\nu}^2 + \tilde y_{\nu}^2 ) - 4 x_e^2 (y_e^2+\tilde y_e^2) -4 x_\nu^2 ( y_{\nu}^2 + \tilde y_{\nu}^2 ) 
-8 (x_l x_e y_e \tilde y_e +  x_l x_\nu y_\nu \tilde y_\nu) ,
  ~~\nonumber  
  \\
 16 \pi^2 \beta_{y_t} & = y_t \left[   \frac{9 y_t^2}{2}+y_e^2+\tilde y_e^2+ y_{\nu}^2 + \tilde y_{\nu}^2   -8 g_3^2 
 - \frac{9 g^2}{4} - \frac{17 g'^2}{12}\right]~~ \nonumber \\
 16 \pi^2 \beta_{y_e} & = y_e \left[ 3 y_t^2 + \frac{5 y_e^2}{2}+\tilde y_e^2 - \frac{y_{\nu}^2}{2} + \tilde y_{\nu}^2  + \frac{x_l^2}{4} + \frac{x_e^2}{4} 
 - \frac{9 g^2}{4} - \frac{15 g'^2}{4}\right]  + x_l x_e \tilde y_e \nonumber \\
  16 \pi^2 \beta_{\tilde y_e} & = \tilde y_e \left[ 3 y_t^2 +  y_e^2+\frac{5 \tilde y_e^2}{2} + y_{\nu}^2 - \frac{\tilde y_{\nu}^2}{2}      
 +\frac{x_l^2}{4} + \frac{x_e^2}{4}   - \frac{9 g^2}{4} - \frac{15 g'^2}{4}\right]  + x_l x_e y_e
 \nonumber \\
 16 \pi^2 \beta_{y_{\nu}} & = y_{\nu} \left[ 3 y_t^2 - \frac{ y_{e}^2}{2}+\tilde y_e^2 + \frac{5 y_{\nu}^2}{2} + \tilde y_{\nu}^2    +\frac{x_l^2}{4} + \frac{x_\nu^2}{4}
 - \frac{9 g^2}{4} - \frac{3 g'^2}{4}\right] +  x_l x_\nu \tilde y_\nu   \nonumber \\
  16 \pi^2 \beta_{\tilde y_{\nu}} & = \tilde y_{\nu} \left[ 3 y_t^2 +  y_e^2-\frac{\tilde y_e^2}{2} + y_{\nu}^2 + \frac{5\tilde y_{\nu}^2}{2}  +\frac{x_l^2}{4} + \frac{x_\nu^2}{4} 
 - \frac{9 g^2}{4} - \frac{3 g'^2}{4} \right] + x_l x_\nu y_\nu  \nonumber \\
  16 \pi^2 \beta_{x_{l}} & = x_{l} \left[ \frac{ y_e^2}{2}+\frac{\tilde y_e^2}{2} + \frac{y_{\nu}^2}{2} + \frac{\tilde y_{\nu}^2}{2}  +\frac{7 x_l^2}{2} + x_e^2+x_\nu^2 
 - \frac{9 g^2}{2} - \frac{3 g'^2}{2} \right] + 2 ( x_e y_e \tilde y_e +  x_\nu y_\nu \tilde y_\nu)  \nonumber \\
  16 \pi^2 \beta_{x_{e}} & = x_{e} \left[  y_e^2+\tilde y_e^2  + 2 x_l^2 + \frac{5 x_e^2}{2} +  x_\nu^2 
 - 6 g'^2 \right] +4 x_e y_e \tilde y_e   \nonumber \\
  16 \pi^2 \beta_{x_{\nu}} & = x_{\nu} \left[  y_\nu^2+\tilde y_\nu^2  + 2 x_l^2 +x_e^2 + \frac{5 x_\nu^2}{2}  \right]  
  +4 x_\nu y_\nu \tilde y_\nu \nonumber \\
 16 \pi^2 \beta_{g_i} & =  g_i^3 b_i,~~~~~~\{ b_{g_3}, b_{g}, b_{g'}\} = \{-7, -\frac{5}{2} ,  \frac{53}{6} \}
 \end{align}

\end{widetext}


\begin{thebibliography}{99}


\bibitem{ATLAS-Higgs} 
  G.~Aad {\it et al.}  [ATLAS Collaboration],
  Phys.\ Lett.\ B {\bf 716}, 1 (2012)
  [arXiv:1207.7214 [hep-ex]].

\bibitem{CMS-Higgs}
  S.~Chatrchyan {\it et al.}  [CMS Collaboration],
  Phys.\ Lett.\ B {\bf 716}, 30 (2012)
  [arXiv:1207.7235 [hep-ex]].
 
\bibitem{:2012zzl} 
  [Tevatron New Physics Higgs Working Group and CDF and D0 Collaborations],
  arXiv:1207.0449 [hep-ex].
   

  
\bibitem{Carena:2011aa} 
  M.~Carena, S.~Gori, N.~R.~Shah and C.~E.~M.~Wagner,
  JHEP {\bf 1203}, 014 (2012)
  [arXiv:1112.3336 [hep-ph]];
   
\bibitem{Batell:2011pz} 
  B.~Batell, S.~Gori and L.~-T.~Wang,
  JHEP {\bf 1206}, 172 (2012)
  [arXiv:1112.5180 [hep-ph]].
   
     
   

  
  
  
  \bibitem{VLfermion}
%
  M.~Carena, I.~Low and C.~E.~M.~Wagner,
  arXiv:1206.1082 [hep-ph];
 %
   N.~Bonne and G.~Moreau,
  arXiv:1206.3360 [hep-ph];
  %
  H.~An, T.~Liu and L.~-T.~Wang,
  arXiv:1207.2473 [hep-ph];
  %
  L.~G.~Almeida, E.~Bertuzzo, P.~A.~N.~Machado and R.~Z.~Funchal,
  arXiv:1207.5254 [hep-ph];
 %
  B.~Batell, D.~McKeen and M.~Pospelov,
  arXiv:1207.6252 [hep-ph];
 %
 J.~Kearney, A.~Pierce and N.~Weiner,
  arXiv:1207.7062 [hep-ph];
%
  H.~Davoudiasl, H.~-S.~Lee and W.~J.~Marciano,
  arXiv:1208.2973 [hep-ph];
%
  K.~J.~Bae, T.~H.~Jung and H.~D.~Kim,
  arXiv:1208.3748 [hep-ph];
  %
  B.~Batell, S.~Gori and L.~-T.~Wang,
  arXiv:1209.6382 [hep-ph].
   

  
\bibitem{Joglekar:2012vc} 
  A.~Joglekar, P.~Schwaller and C.~E.~M.~Wagner,
  arXiv:1207.4235 [hep-ph].
  
\bibitem{ArkaniHamed:2012kq} 
  N.~Arkani-Hamed, K.~Blum, R.~T.~D'Agnolo and J.~Fan,
  arXiv:1207.4482 [hep-ph].
  
\bibitem{Reece:2012gi} 
  M.~Reece,
  arXiv:1208.1765 [hep-ph].


\bibitem{lpp2}

  H.~M.~Lee, M.~Park and W.~-I.~Park,
  arXiv:1209.1955 [hep-ph].


  
 
  
\bibitem{EliasMiro:2012ay} 
  J.~Elias-Miro, J.~R.~Espinosa, G.~F.~Giudice, H.~M.~Lee and A.~Strumia,
  JHEP {\bf 1206}, 031 (2012)
  [arXiv:1203.0237 [hep-ph]].
  
           

\bibitem{lebedev}   
  O.~Lebedev,
  Eur.\ Phys.\ J.\ C {\bf 72}, 2058 (2012)
  [arXiv:1203.0156 [hep-ph]].

 



  

\bibitem{Dobrescu:2000jt} 
  B.~A.~Dobrescu, G.~L.~Landsberg and K.~T.~Matchev,
  Phys.\ Rev.\ D {\bf 63}, 075003 (2001)
  [hep-ph/0005308].

\bibitem{Draper:2012xt} 
  P.~Draper and D.~McKeen,
  Phys.\ Rev.\ D {\bf 85}, 115023 (2012)
  [arXiv:1204.1061 [hep-ph]].

\bibitem{Toro:2012sv} 
  N.~Toro and I.~Yavin,
  Phys.\ Rev.\ D {\bf 86}, 055005 (2012)
  [arXiv:1202.6377 [hep-ph]].

\bibitem{Ellis:2012sd} 
  S.~D.~Ellis, T.~S.~Roy and J.~Scholtz,
  arXiv:1210.1855 [hep-ph];
  S.~D.~Ellis, T.~S.~Roy and J.~Scholtz,
  arXiv:1210.3657 [hep-ph].

    
    
    
  
\bibitem{Ellis:1975ap} 
  J.~R.~Ellis, M.~K.~Gaillard and D.~V.~Nanopoulos,
  Nucl.\ Phys.\ B {\bf 106}, 292 (1976).
 
  
\bibitem{Shifman:1979eb} 
  M.~A.~Shifman, A.~I.~Vainshtein, M.~B.~Voloshin and V.~I.~Zakharov,
  Sov.\ J.\ Nucl.\ Phys.\  {\bf 30}, 711 (1979)
  [Yad.\ Fiz.\  {\bf 30}, 1368 (1979)].
  
   
   
   
\bibitem{Isidori:2001bm} 
  G.~Isidori, G.~Ridolfi and A.~Strumia,
  Nucl.\ Phys.\ B {\bf 609}, 387 (2001)
  [hep-ph/0104016].



  



   
\bibitem{Voloshin:2012tv} 
  M.~B.~Voloshin,
  arXiv:1208.4303 [hep-ph].
   
\bibitem{McKeen:2012av} 
  D.~McKeen, M.~Pospelov and A.~Ritz,
  arXiv:1208.4597 [hep-ph].



 
 \bibitem{pko}
  S.~Baek, P.~Ko, W.~-I.~Park and E.~Senaha,
  arXiv:1209.4163 [hep-ph].



\bibitem{Djouadi:2005gi} 
  A.~Djouadi,
  Phys.\ Rept.\  {\bf 457}, 1 (2008)
  [hep-ph/0503172].




  
 
\bibitem{singlets} 
  C.~Englert, T.~Plehn, M.~Rauch, D.~Zerwas and P.~M.~Zerwas,
  Phys.\ Lett.\ B {\bf 707}, 512 (2012)
  [arXiv:1112.3007 [hep-ph]];
  I.~Low, J.~Lykken and G.~Shaughnessy,
  arXiv:1207.1093 [hep-ph];
   %
   D.~Carmi, A.~Falkowski, E.~Kuflik, T.~Volansky and J.~Zupan,
  arXiv:1207.1718 [hep-ph];
  %
  D.~Bertolini and M.~McCullough,
  arXiv:1207.4209 [hep-ph];
  B.~Batell, D.~McKeen and M.~Pospelov,
  JHEP {\bf 1210}, 104 (2012)
  [arXiv:1207.6252 [hep-ph]].
  


  
  

\bibitem{Peskin:1991sw} 
  M.~E.~Peskin and T.~Takeuchi,
  Phys.\ Rev.\ D {\bf 46}, 381 (1992).




\bibitem{Baak:2012kk} 
  M.~Baak, M.~Goebel, J.~Haller, A.~Hoecker, D.~Kennedy, R.~Kogler, K.~Moenig and M.~Schott {\it et al.},
  arXiv:1209.2716 [hep-ph].
  


\bibitem{singletpheno} V.~Silveira and A.~Zee,
  Phys.\ Lett.\ B {\bf 161}, 136 (1985); T.~Binoth and J.~J.~van der Bij,
  Z.\ Phys.\ C {\bf 75}, 17 (1997)
  [hep-ph/9608245]; J.~McDonald,
  Phys.\ Rev.\ D {\bf 50}, 3637 (1994)
  [hep-ph/0702143 [HEP-PH]]; C.~P.~Burgess, M.~Pospelov and T.~ter Veldhuis,
  Nucl.\ Phys.\ B {\bf 619}, 709 (2001)
  [hep-ph/0011335]; 
  R.~Schabinger and J.~D.~Wells,
  Phys.\ Rev.\ D {\bf 72}, 093007 (2005)
  [hep-ph/0509209];
  D.~O'Connell, M.~J.~Ramsey-Musolf and M.~B.~Wise,
  Phys.\ Rev.\ D {\bf 75}, 037701 (2007)
  [hep-ph/0611014];
  M.~Bowen, Y.~Cui and J.~D.~Wells,
  JHEP {\bf 0703}, 036 (2007)
  [hep-ph/0701035];
V.~Barger, P.~Langacker, M.~McCaskey, M.~J.~Ramsey-Musolf and G.~Shaughnessy,
  Phys.\ Rev.\ D {\bf 77}, 035005 (2008)
  [arXiv:0706.4311 [hep-ph]];
  S.~Gopalakrishna, S.~Jung and J.~D.~Wells,
  Phys.\ Rev.\ D {\bf 78}, 055002 (2008)
  [arXiv:0801.3456 [hep-ph]];
    P.~J.~Fox, D.~Tucker-Smith and N.~Weiner,
  JHEP {\bf 1106}, 127 (2011)
  [arXiv:1104.5450 [hep-ph]];  
  I.~Low, J.~Lykken and G.~Shaughnessy,
  Phys.\ Rev.\ D {\bf 84}, 035027 (2011)
  [arXiv:1105.4587 [hep-ph]];
   M.~J.~Dolan, C.~Englert and M.~Spannowsky,
  arXiv:1210.8166 [hep-ph].




\bibitem{lpp1}

  H.~M.~Lee, M.~Park and W.~-I.~Park,
  arXiv:1205.4675 [hep-ph].




\bibitem{2HDM} 
  P.~M.~Ferreira, R.~Santos, M.~Sher and J.~P.~Silva,
  Phys.\ Rev.\ D {\bf 85}, 077703 (2012)
  [arXiv:1112.3277 [hep-ph]];
  K.~Blum and R.~T.~D'Agnolo,
  Phys.\ Lett.\ B {\bf 714}, 66 (2012)
  [arXiv:1202.2364 [hep-ph]];
  A.~Azatov, S.~Chang, N.~Craig and J.~Galloway,
  Phys.\ Rev.\ D {\bf 86}, 075033 (2012)
  [arXiv:1206.1058 [hep-ph]];
  N.~Craig and S.~Thomas,
  arXiv:1207.4835 [hep-ph];
  D.~S.~M.~Alves, P.~J.~Fox and N.~J.~Weiner,
  arXiv:1207.5499 [hep-ph];
  W.~Altmannshofer, S.~Gori and G.~D.~Kribs,
  arXiv:1210.2465 [hep-ph].
   


\bibitem{RG} 
  M.~E.~Machacek and M.~T.~Vaughn,
  Nucl.\ Phys.\ B {\bf 222}, 83 (1983);
  M.~E.~Machacek and M.~T.~Vaughn,
  Nucl.\ Phys.\ B {\bf 236}, 221 (1984);
  M.~E.~Machacek and M.~T.~Vaughn,
  Nucl.\ Phys.\ B {\bf 249}, 70 (1985);
   M.~-x.~Luo, H.~-w.~Wang and Y.~Xiao,
  Phys.\ Rev.\ D {\bf 67}, 065019 (2003)
  [hep-ph/0211440].


\end{thebibliography}
\end{document}